# The Sustainability Paradox of Biodegradable Packaging Through a Life Cycle Perspective on Chitosan Based Food Packaging

Maria Cosic[1], Milena Cosic[1], Amy Kim[2], Siddhartha Pahari[1], Grace Wu[1], Rayna Zhou[1], Trisha Daftari[3], Shaily Gupta[4], Reyna Michelle Suneel[5], Srinivasan Latha[6, *], Rashi Sharma[7], Utkarsh Chadha[8, 9, *]

[1]Department of Chemical Engineering and Applied Chemistry, Faculty of Applied Sciences and Engineering, **University of Toronto**, Toronto, ON, M5S 3E4, Canada

[2]Department of Civil and Mineral Engineering, Faculty of Applied Sciences and Engineering, **University of Toronto**, Toronto, Ontario ON M5S 1A4, Canada

[3]College of Engineering, **Georgia Institute of Technology**, North Avenue Atlanta, GA 30332

[4]Harold and Inge Marcus Department of Industrial and Manufacturing Engineering, **Pennsylvania State University**, 310 Leonhard Building, University Park, PA 16802, United States

[5]**University of Leeds**, Leeds LS2 9JT, United Kingdom

[6]Department of Chemistry, School of Advanced Sciences (SAS), **Vellore Institute of Technology**, Vellore, Tamil Nadu 632014, India

[7]Department of Biomedical Engineering, College of Engineering, **Texas A&M University**, College Station, TX 77801, USA

[8]Department of Mechanical and Industrial Engineering, Faculty of Applied Sciences and Engineering, **University of Toronto**, Toronto, Ontario ON M5S 3G8, Canada

[9]Department of Materials Science and Engineering, Faculty of Applied Sciences and Engineering, **University of Toronto**, Toronto, Ontario ON M5S 3E4, Canada

*Corresponding authors: srinivasan.latha@vit.ac.in (S. Latha); utkarshchadha1302@gmail.com (U. Chadha);

## Abstract

Chitosan-based packaging systems are being studied as alternatives to traditional food packaging made from petroleum sources. The research that has been carried out so far has mostly focused on analyzing their material properties, formulation strategies, or biodegradable behavior. However, as a result, there is a gap left in understanding how the upstream impacts affect the functional advantages of food packaging and the end-of-life results of the formulation process. This review evaluates what trade-offs exist as well as what factors affect the sustainability of chitosan-based packages. Beginning with chitin extraction from crustacean waste, key processing steps, including demineralization, deproteinization, and nanoparticle synthesis, are assessed in terms of chemical intensity, energy demand, and associated emissions. Manufacturing routes, including solvent-based and green synthesis approaches, are compared with those of conventional plastics to identify relative environmental burdens. The use phase is analyzed with respect to antimicrobial functionality, shelf life extension, and potential reductions in food waste. End-of-life pathways, including biodegradation and composting, are evaluated alongside uncertainties related to degradation behavior and nanoparticle fate. The review takes into account life-cycle issues and offers a scientifically based viewpoint on how to differentiate between packaging applications whose environmental benefits can be determined and those whose ecological footprint, lack of functionality, and uncertainty regarding disposal make it impossible to assess sustainability.

---

## 1. Introduction

Plastic packaging has become indispensable in modern food supply chains because of its lightweight nature, barrier performance, and ability to preserve food quality during storage and transportation [1-4]. However, the environmental concerns associated with single-use plastics have increased because of persistent waste accumulation, microplastic contamination, and increasing pressure on waste management systems [5-6]. Global plastic production has exceeded 390 million tonnes annually, as reported by Plastics Europe in their report in 2023 [7], where the packaging industry represented the largest application sector for plastics, accounting for nearly 39% of total European plastic demand [7]. A large proportion of plastic packaging is designed for short service lifetimes and is frequently discarded within months of production, contributing significantly to global waste generation and environmental leakage [8]. Furthermore, mismanaged plastic waste has resulted in significant environmental leakage, with several million tonnes estimated to enter marine environments each year [6-8].

In response to these concerns, increasing attention has been given to biobased and biodegradable materials as alternatives to petroleum-derived polymers. Among these candidates, polysaccharide-based materials such as chitosan have attracted considerable interest because of their renewable origin, intrinsic antimicrobial activity, and film-forming capability [9]. Chitosan, derived primarily from the deacetylation of chitin present in crustacean shells and other marine biomasses, has been widely investigated for food packaging applications where its oxygen barrier properties and antimicrobial functionality can contribute to improved food preservation [10]. Recent research has further explored the incorporation of nanomaterials and functional additives to increase the mechanical strength, moisture resistance, and shelf-life performance of chitosan-based packaging systems [11-13].

Recent literature estimates that approximately 6–8 million tonnes of crustacean shell waste from crab, shrimp, and lobster processing are generated globally each year, with a substantial proportion still being discarded [14]. This large and continuing waste stream provides an important source of chitin-rich biomass for valorization into chitosan and other value-added materials. The tonnage of waste from crustaceans alone tends to increase over time, and its disposal is both costly and contributes to marine environmental pollution [15]. Therefore, the abundance of this waste presents a significant opportunity for the use of chitin derivatives in food packaging applications. Since chitin is the second most abundant natural polysaccharide found within the shells of crustacean waste, the use of these biobased wastes from the seafood industry can help reduce the reliance on energy-intensive fossil fuels to produce the conventional plastic films commonly utilized in the market. Within crustacean shell waste, the levels of chitin constituents vary depending on species, distance of collection, and processing type. According to a recent review, the chitin content of crustacean shell residues is between 15% and 40%, depending on how it was evaluated and calculated [216]. Hence, one cannot generalize the data obtained regarding the chitin constituent for all types of waste from the crustacean industry [216].

While such developments are frequently positioned within the broader transition toward sustainable materials, the environmental implications of biobased packaging cannot be fully understood by considering material origin or biodegradability alone [17, 18]. In many cases, the sustainability narrative surrounding biopolymers emphasizes renewable feedstocks and end-of-life degradability, whereas less attention is given to the environmental burdens associated with extraction, chemical processing, modification, and large-scale manufacturing [19, 20]. For instance, conventional chitosan production typically involves multiple acid and alkaline treatments, high water consumption, and thermal processing steps that can generate chemical effluents and energy demands. These upstream impacts raise the important question of whether the environmental advantages attributed to biobased materials remain valid when they are evaluated across their entire life cycle.

### *1.1. Review Objectives*

A central question addressed in this review is the sustainability paradox associated with bio-based food packaging. Materials derived from renewable resources or designed for biodegradability are often described as environmentally favourable alternatives to conventional plastics, yet these claims can be weakened by upstream chemical processing, energy demand, wastewater generation, additive-related uncertainty, and infrastructure-dependent end-of-life behaviour. In chitosan-based systems, these tensions are especially important because the material is frequently positioned as both waste-derived and biodegradable, even though its production commonly depends on acid–alkali extraction, multiple washing steps, and formulation strategies that can alter degradation pathways and food-contact safety.

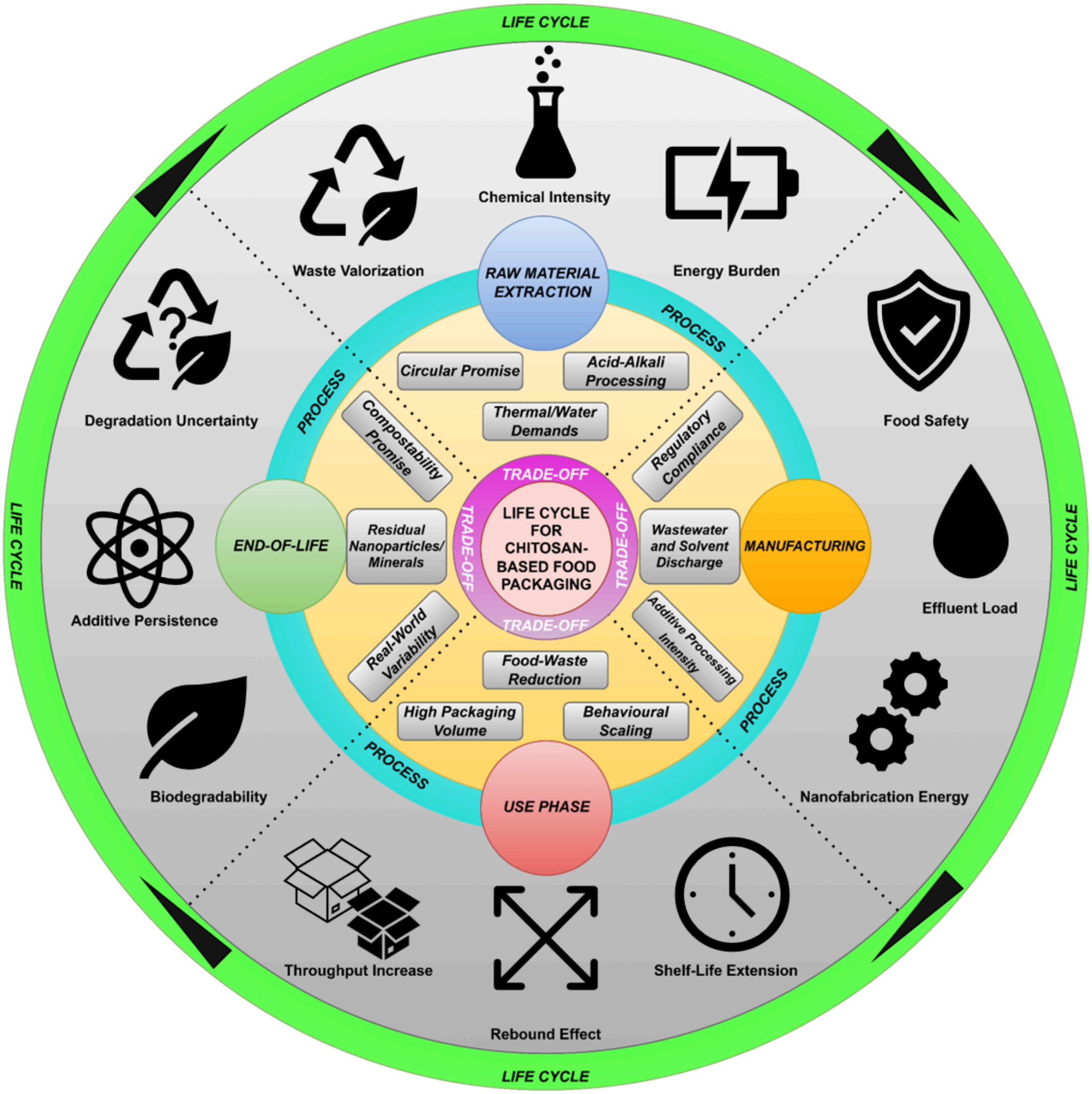

**Figure 1. Life cycle trade-off taxonomy for chitosan-based food packaging systems.** *The figure presents a conceptual mapping of stagewise interactions across raw material extraction, manufacturing, use phase, and end-of-life, highlighting key processing pathways and associated sustainability trade-offs. The upstream stages are characterized by chemical intensity, energy burden, and wastewater generation, whereas the use phase reflects functional benefits such as*

*shelf life extension and food waste reduction, in addition to potential rebound effects. End-of-life outcomes emphasize biodegradability alongside uncertainties related to additive persistence and degradation pathways. The framework illustrates the uneven distribution of environmental burdens and functional gains across the life cycle, challenging simplified sustainability narratives based solely on biodegradability.*

The connection between the material used and the production processes determines how environmentally friendly chitosan used for food packaging is, as shown in Figure 1. This schematic provides an overview of the interconnected life-cycle stages and their potential trade-offs; the evidence-based interpretation of the principal environmental drivers is further presented in Figure 6. It is evident that certain processes during the preparation of raw materials and the production of chitosan such as acid-alkali extraction, washing, and deacetylation lead to chemical consumption, water consumption and energy needs depending on how efficiently the extraction processes, the recovery of chemicals, the conditions of production and energy supply system work. The impact of these processes must also be compared to the measurable advantages of using that packaging, namely barrier properties of chitosan packaging, the antimicrobial effect of chitosan and shelf-life of products that can be maintained using chitosan packages.

Accordingly, this review adopts a life-cycle trade-off perspective rather than a simplified substitution narrative. It does not present a new quantitative life-cycle assessment. Instead, it synthesizes evidence from materials science, packaging research, environmental assessment, and food-contact literature to identify where environmental gains are robust, where burdens are shifted between life-cycle stages, and which variables most strongly determine the overall sustainability of chitosan-based food packaging. By consolidating these fragmented perspectives, the review provides a structured basis for evaluating chitosan beyond claims of biodegradability or renewable origin alone. The key contribution of this review is to combine three related topics that are usually studied separately: environmental impact of chitosan production, functional properties and preservation advantages of modified packaging solutions, and uncertainties related to disposal of modified packaging. The holistic perspective presented here allows for identifying the issue of shifting environmental burden and determining when, in what circumstances chitosan-based packaging could outperform alternatives that are functionally similar. In this review, "chitosan-based packaging" is used as an umbrella term encompassing neat chitosan films and coatings, blends, and modified formulations; "chitosan nanocomposite" or "nano-enabled chitosan packaging" is used only where nanoscale components are specifically present.

### *1.2. Methodology*

#### 1.2.1. Research Questions

This review is guided by the following research questions, which aim to examine the environmental implications of chitosan-based food packaging from a life cycle perspective:

- *What are the major environmental considerations associated with the raw material sourcing and processing of the chitosan used in food packaging applications?*
- *How do manufacturing processes and additive incorporation influence the environmental profile of chitosan-based packaging systems?*
- *To what extent can functional benefits during the use phase, such as shelf life extension or antimicrobial performance, contribute to environmental gains such as reduced food waste?*
- *What uncertainties and trade-offs arise at the end-of-life stage, particularly regarding biodegradability claims and the environmental fate of additives?*
- *How do these stage-specific impacts interact across the life cycle, and what implications do they have for evaluating the sustainability of chitosan-based packaging materials?*

### 1.2.2. Search Strategy

To address these research questions, a narrative literature review approach was adopted. Relevant publications were identified using academic databases, including Scopus, Web of Science, and Google Scholar, with a focus on studies related to chitosan extraction, processing methods, nanocomposite packaging materials, environmental impacts, and life cycle considerations. Keywords included combinations of *"chitosan," "biopolymer packaging," "chitosan nanocomposites," "food packaging sustainability,"* and *"life cycle impacts."*

Both peer-reviewed articles and selected reports/gray literature sources from international organizations were considered to capture broader environmental and industrial perspectives, as expanded upon in Section 1.2.3. A systematic review with PRISMA protocol was not undertaken because the research questions span multiple disciplines (materials synthesis, food packaging performance, and life-cycle trade-offs) and because high-quality, full life-cycle inventory data for chitosan-nanocomposite packaging systems remain too limited for meaningful quantitative meta-analysis.

In the life-cycle assessment, the studies published were analyzed based on their functional units, system boundaries, environmental indicators, and modeling assumptions. The assessments from the production-stage assessments were expressed per kilogram of chitosan, compared with film assessments expressed per square metre, and even more general food packaging analyses based on certain preservation functions. The system boundaries from each study were examined separately and included cradle-to-grave assessments and studies that cover manufacturing, usage, and end-of-life processes. The interpretation made took into consideration the variances in feedstock treatment and allocation/substitution of raw materials, energy supply, and process inventories. As this is a narrative literature review and not an original LCA in its quantitative form, there was no need to create a new life-cycle inventory or a harmonized impact-assessment model. Therefore, the numerical results from different papers can't be compared directly unless they are similar enough in their methodology, assumptions, and functional units used.

### 1.2.3. Evidence Transparency

To ensure transparency in the use of evidence, Table 1 summarizes the source types included in this narrative-conceptual review and clarifies their role in the analysis. Peer-reviewed literature was prioritized for technical claims related to chitosan properties, synthesis routes, film performance, biodegradation, nanomaterial behavior, and life cycle impacts. Standards and institutional documents were used to support methodological, regulatory, and policy-related discussion. Industry reports, technical documentation, and trade or news sources were used only for contextual information, such as market trends, industrial practices, or examples not extensively covered in peer-reviewed literature.

The evidence base was classified into six categories: peer-reviewed journal articles, conference proceedings, standards and institutional documents, industry/market reports and company documentation, news or trade publications, and books, academic book chapters, and theses. This classification was used to distinguish analytical evidence from contextual evidence and to avoid treating all source types with equal evidentiary weight. Accordingly, peer-reviewed and standards-based sources formed the primary basis for scientific interpretation, whereas grey literature was used selectively to support broader industrial and practical context.

In assessing the evaluated literature, it is important to distinguish between results based on experimental measurements, results derived from theoretical modeling, and results that are not verified to apply to real-world conditions. Results obtained experimentally were interpreted based upon their experimental conditions, while LCA results were judged only in terms of assumptions made and system boundaries applied in the original study. In cases where there was no formulation-specific data on migration, environmental persistence, ecotoxicity, and ultimate biodegradation the result was treated as uncertainty instead of an established effect.

**Table 1.** Classification of sources used in this review.

| Source Types | Role in Analysis | Primary Sections Used | Evidence Priority | Number of Sources Cited in References (n) | Percentage of Total |
|---|---|---|---|---|---|
| Peer-reviewed articles | Core scientific evidence, material properties, processing routes, environmental impacts, packaging performance, LCA findings | Sections 2–6 (Chitosan Nanomaterials; Life cycle interpretation; Understanding Trade-offs; Comparative Analysis; Current Drawbacks and | Primary | 198 | 88.79% |
| | | Challenges), with supporting use in Section 7 (Outlook and Recommendations) | | | |
| Conference proceedings | Supplemental technical developments and emerging technologies | Sections 2.2 and 3.2–3.3 (Synthesis Methods; Manufacturing pathway choices; Additive selection | Primary | 2 | 0.9% |
| Standards & institutional documents | Regulatory guidance, food-contact requirements, environmental frameworks, production statistics | Sections 1.2.3, 3.1–3.2, and 5.4–5.6 (Evidence transparency; Raw material origin; Manufacturing pathway choices; Functional performance; Use phase; End-of-life realism) | Secondary | 7 | 3.14% |
| Industry reports & company documentation | Market trends, industrial production, commercialization context | Sections 1, 5, and 7 (Introduction; Comparative Analysis; Outlook and Recommendations | Contextual | 4 | 1.79% |
| News articles/Trade Publications | Background context and industry developments | Sections 3–5 (Life cycle interpretation; Trade-offs; Comparative Analysis) | Background, Contextual | 1 | 0.45% |
| Books, Academic Textbooks & Theses | Foundational concepts, packaging science, benchmark material properties, extraction chemistry, and supporting background information | Sections 2, 3, and 5 (Functional role; Life cycle interpretation; Comparative Analysis) | Secondary | 11 | 4.93% |

| | | |
|---|---|---|
| **Peer-reviewed scientific literature (articles + conference proceedings)** | **200** | **89.69%** |
| **Grey literature (standards/institutional + industry + news/trade + Books/Textbooks/Theses)** | **23** | **10.31%** |
| **Total References** | **223** | **100%** |

## 2. Chitosan Based Materials and Nanocomposites in Food Packaging.

Research has been conducted on the use of chitosan in the manufacture of food-related films and coatings due to its film-forming capabilities, antimicrobial action, and oxygen-barrier properties in optimal humidity conditions. Nevertheless, functionality or the performance of chitosan packaging materials is governed by polymer properties, composition, film thickness, and ambient conditions. Consequently, all of these factors determine the processing needs for using a certain formulation, and in turn, decide whether this formulation can be compared in terms of life cycle. The following subsections expand on the basic properties of chitosan, synthesis processes, and food packaging-specific details.

### *2.1. Properties Facilitating Food Packaging Application*

The primary amino groups of chitosan exhibit an apparent pKa of approximately 6.3–6.5. At pH values below this range, protonation of $–NH_2$ groups to $–NH_3^+$ increases the cationic character, hydration, and solubility of chitosan, whereas progressive deprotonation at higher pH reduces these effects. A higher degree of deacetylation generally increases the number of available amino groups and therefore the potential cationic charge density under acidic conditions; however, the apparent pKa and resulting film behaviour are also influenced by molecular weight, acetyl-group distribution, ionic environment, crystallinity, and formulation. Consequently, no single degree-of-deacetylation threshold can universally predict swelling or degradation behaviour. In acidic food environments, increased protonation can promote hydration, swelling, and partial dissolution of chitosan films, with the extent of these effects further controlled by molecular weight, crosslinking, crystallinity, and incorporated additives [217, 218].

The biopolymer Chitosan is produced by alkaline N-deacetylation of chitin isolated from crustacean shells. Chitosan can be used as packaging material for food because it has cationic amino functional groups, hydroxyl functional groups, crystalline structure, tunable molecular weight and degree of de-acetylation. These factors affect the antimicrobial effectiveness, film-forming properties, mechanical properties and barrier performance of chitosan in food packaging applications [21-24]. There are a number of commercial chitosan grades that differ in molecular weight and degree of deacetylation (DD). These parameters influence solubility, charge density, viscosity, chain packing, and interaction with additives, which are crucial to the performance of the film as a packaging material. Figure 2 illustrates the chemical structure and major functional modification sites of chitosan [215].

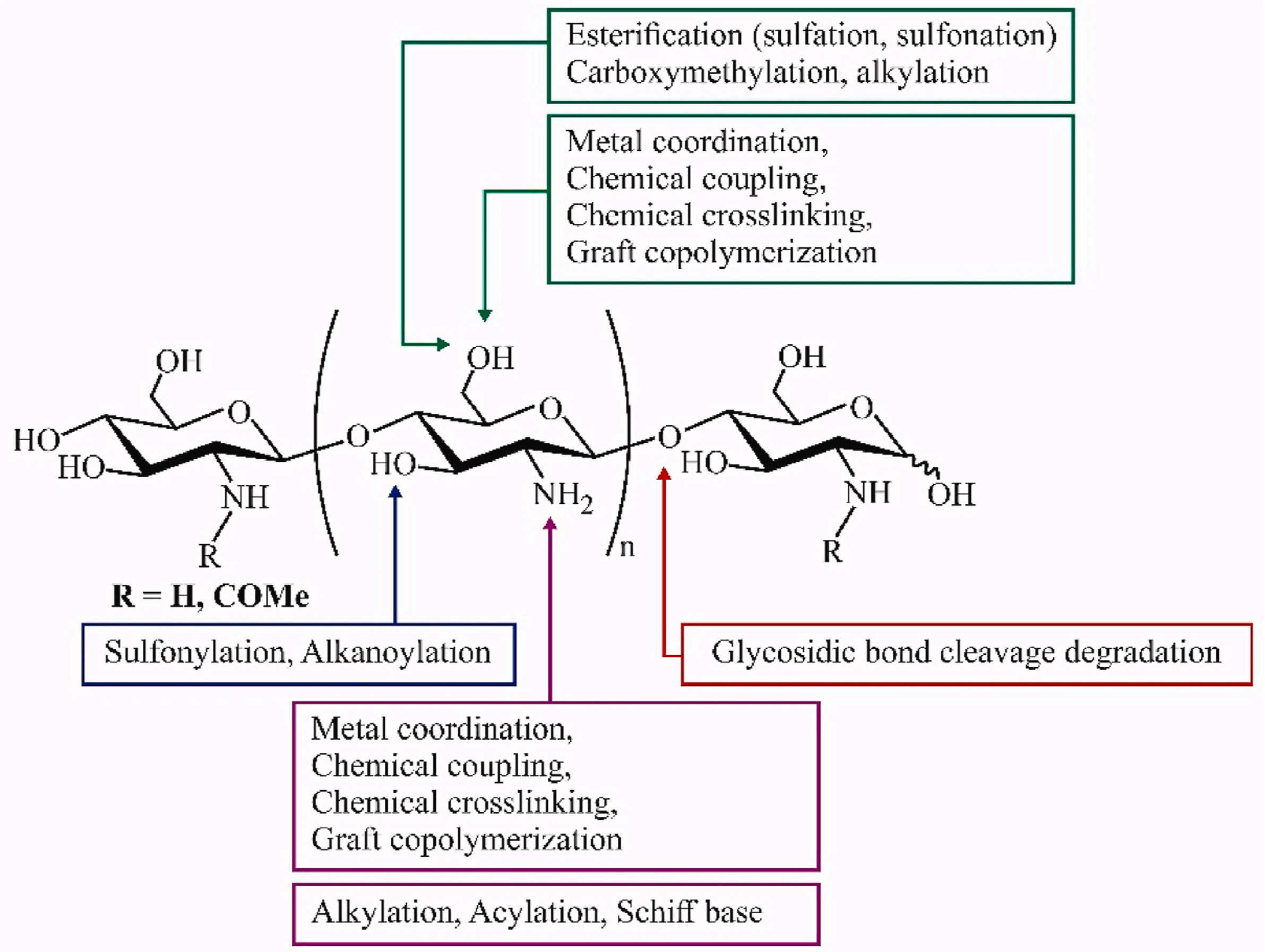

**Figure 2.** The relationship between DA and DD. The repeating unit on the left is N-acetyl-D-glucosamine, while the one on the right is β-1,4-D-glucosamine. **Reprinted from Edo et al. [215], *Carbohydrate Research*, Vol. 550, Article 109409, Copyright © 2025 Elsevier B.V., with permission from Elsevier.**

Chitosan films exhibit a considerable variance in their mechanical properties based upon the molecular weight, the degree of deacetylation, and how they are formulated with additives. Medium to high molecular weight chitosan typically can provide a useful tensile strength, and in conjunction with other materials such as plasticizers, PVA, silica and, natural extracts are commonly added to produce improved flexibility, transparency and handling properties. However, chitosan alone usually has a relatively low elongation and may need to be modified from its original formulation to be used in packaging applications [25-29].

Chitosan's antimicrobial activity is associated partly with its protonatable amino groups, which enable electrostatic interactions with negatively charged microbial cell surfaces. Its positive charges interact electrostatically with negatively charged microbial cell membranes. These interactions can damage microbial cell membranes, resulting in increased permeability and cell death. This is particularly effective against a broad spectrum of bacteria, both gram-positive and gram-negative [30]. Notably, conditions of neutral or alkaline pH often do not promote such properties [31].

Chitosan can also disrupt the structures of microbial cell membranes. It might bind to microbial DNA after it enters the cell, preventing transcription and translation from occurring. This interference can control the synthesis of proteins, preventing the development and reproduction of microorganisms. Through membrane disruption and DNA binding as one of the mechanisms, chitosan functions as a powerful antibacterial agent [32]. Chitosan films possess inherent antimicrobial and antioxidant capabilities in contrast to traditional plastics that are solely dependent on the addition of active agents. Chitosan is typically an effective oxygen barrier under dry conditions, but its hydrophilicity leads to high water uptake and weak water-vapor barrier performance under humid conditions unless modified by blending, crosslinking, hydrophobic additives, or multilayer design [28, 33].

Furthermore, the stiff crystalline structure of chitosan provides a strong oxygen barrier, especially when it is dry. On the other hand, because chitosan is hydrophilic, its ability to act as a water vapor barrier is moderate [34]. Consequently, moisture absorption can cause chitosan films to swell and increase water-vapor transmission, with the extent of these effects depending on polymer characteristics and formulation conditions. A higher molecular weight and less deacetylation can lead to a lower water permeability and vice versa.

On the other hand, on its own, the water permeability of chitosan is far less than that of water vapor plastics such as LDPE, HDPE, and PP [28]. Additives can enhance this resistance, and the ability to change the moisture permeability allows the material properties to be tailored to different needs within food packaging. In terms of gases, compared with commercially used plastics, chitosan film has a low oxygen permeability, resulting in a longer shelf life and a slower oxidation rate (i.e., lower rancidity) [22]. Both a strong oxygen barrier and metal ion chelation prevent lipid oxidation and are considered the main mechanisms underlying the antioxidant properties of chitosan. Chitosan coatings and films can lower the oxidation rate to 0.08 to 31.07 $cm^3$ $m/mm^2$.day.atm [35].

In comparison to many commercial plastics, chitosan has a lower oxygen permeability because of its high crystallinity and high number of intra- and inter-molecular interactions [36]. While each additive has its own property, it also has its own drawbacks. Chitosan itself, while considered a material that can be used for a diverse amount of food packaging, is limited by its rapid dissolution in acidic solutions and low thermal stability. Preserving acidic food in chitosan films or coatings could affect not only the quality of the films but also the quality of the food.

### 2.2. Synthesis Methods

The use of chitosan-based films and coatings, to an extent, delays food spoilage, protects against microbial contamination, and thereby extends food shelf life. The common methods employed for synthesizing chitosan-based films are briefly discussed below and expanded upon in **Table 2.** Solvent casting, the most widely utilized synthesis method, involves the dissolution of chitosan within dilute organic acids, such as acetic acid. This creates a homogenous chitosan solution that is cast on a flat surface that forms a thin film when the solvent evaporates [37]. This approach allows for exact control over the thickness and qualities of chitosan-based films, making it an easy-to-use and reasonably affordable process [37].

Coating-based techniques, including dipping, spray coating and electrospraying, are utilized to coat various substrates. In the dipping and spraying method, a substrate is applied for an allotted time, either by spraying or dipping, after which the chitosan solution is withdrawn and dried [38]. Both methods are relatively simple with easy control over the film coating thickness, although compared with other film synthesis methods, both methods may result in lower mechanical strength and barrier properties [38]. Electrospraying involves exposing a chitosan solution to increasing electric fields that induce the splitting of charged liquid spray into micro/nanodroplets that are then deposited on a collector, enabling the formation of thin layers.

The layer-by-layer method (LbL) is a versatile approach that allows precise control over the thickness and properties of multilayer films. This enables the incorporation of different materials to form films that exhibit excellent mechanical strength, stability, and controlled release properties [39]. This is a repetitive process in which successive layers of chitosan and other materials are deposited onto a flat-surfaced substrate by immersing them into a solution containing the desired material and then rinsing them to remove any excess material [40]. Solution blending is often utilized to enhance the functional performance of chitosan-based films. A homogeneous chitosan solution is blended with other materials, often with polymers/PVA/starch/gelatin/nanofillers, and poured onto a flat surface, spread out, and dried to create a solid film. [41].

The extrusion process begins with plasticization and blending (often with thermoplastics/biopolymers) of chitosan to enable melt processability. It is mixed to form a homogeneous blend and fed into an extruder, which rotates and shapes the blend into a continuous film [38]. Chitosan has limited thermal stability; therefore, high temperatures may chemically alter the properties of chitosan films, such as by decreasing their mechanical strength. [42-44]. Chitosan nanoparticles can be produced using ionic gelation through the electrostatic binding of the protonated form of chitosan ($–NH_3^+$) with a variety of multivalent anionic agents (usually tripolyphosphate (TPP)). To produce nanoparticles, the anionic cross-linker is added in small increments to a chitosan solution under a controlled pH and agitation and then stabilized and purified. [45] Both the size of the nanoparticles and their surface charge influence with food-packaging-relevant endpoints such as migration, release behavior, dispersion stability, food-contact safety and biocompatibility of the produced film. In addition, this process enables efficient encapsulation of active agents within the chitosan matrix; however, batch-to-batch variance is possible since external factors, such as the pH and concentration of solutions, can easily affect the results [45].

The emulsion-based process used to create chitosan films utilizes mechanical homogenization to produce fine droplets, while stabilizers are added to prevent droplet coalescence. Without stabilization, emulsions may experience phase separation and loss of emulsion properties that contribute to determining the final product's shelf life and performance [46, 47]. Electrospinning, or commonly known as electrostatic whirling, is an efficient method to produce ultrafine fibers with diameters ranging from nanometers to microns. This process is highly sensitive to environmental humidity and is performed by applying a strong electric field to a homogeneous polymer solution [48, 49].

In this method, 2% (v/v) aqueous acetic acid is dissolved in chitosan and blended with poly(vinyl alcohol) to improve electrospinnability. Under high voltage, a Taylor cone forms at a metal capillary tip, resulting in the ejection of a polymer jet solidifying into the fibers. The solvent evaporates when the jet moves closer to the grounded collector [50]. The resulting electrospun chitosan/PVA membranes have a high surface area and porosity, making them suitable for use in a variety of industrial applications [51]. In this context, green synthesis refers to strategies employed for preparing and processing materials aimed at minimising the use of harmful reagents, solvents, energy consumption, and waste produced while achieving the desired quality and functionality of the product. The mere use of renewable raw materials or various solvents does not imply less harmful environmental effects, and this has to be verified by comparing the amount of raw materials and energy spent on the processes and the resulting yields as well as the need for waste disposal [52, 53]. Catalysis is a popular method in green synthesis because it speeds up chemical reactions and results in the generation of less waste. Reusable catalysts are frequently used in catalytic processes to reduce resource consumption and increase efficiency [54-57].

**Table 2.** Synthesis Methods for Chitosan-based Film Materials for Food Packaging-related Applications

| **Process or Method** | **Typical Materials Used** | **Key Processing Parameters** | **Main Advantages** | **Main Disadvantages** | **Typical Applications** | **Ref** |
|---|---|---|---|---|---|---|
| **Solvent casting** | Chitosan (solute) dissolved in weak acid (like acetic, lactic, or citric acid) + optional plasticizers, crosslinkers, or fillers | - Dissolve chitosan in acid until fully solubilized.<br>- Degas to remove bubbles.<br>- Pour onto a leveled plate or mold and dry at controlled temperature and humidity until the solvent evaporates. | - Simple and easy to scale up in the lab.<br>- Good control over film thickness and uniformity.<br>- Compatible with many additives (nanoparticles, essential oils, proteins, etc.).<br>- Produces continuous, defect-free films with a good oxygen barrier. | - Drying is time-consuming.<br>- Mechanical strength can be low and films may be brittle without plasticizer.<br>- Barriers to water vapor are often limited due to hydrophilicity.<br>- Residual solvent and external factors (temperature, airflow) affect properties. | - Lab-scale development of active and intelligent films.<br>- Multilayer structures combining chitosan with other biopolymers. | [37] |
| **Dipping (dip coating)** | Plastic/metal/glass substrate + chitosan solution; may include plasticizers, nanoparticles, or bioactives | - Control chitosan concentration and viscosity.<br>- Immerse clean substrate for defined time, withdraw at constant speed.<br>- Allow draining and drying under controlled temperature/humidity.<br>- Repeat cycles to increase coating thickness. | - Very easy to perform with simple equipment.<br>- Good control over coating thickness through concentration and number of dips.<br>- Can coat complex shapes (fruits, meat pieces, trays). | - Wastage of chitosan solution and poor material efficiency.<br>- Low mechanical strength of stand-alone coating.<br>- Non-uniformity at edges or complex geometries.<br>- Lower barrier to water vapor compared to dense cast films. | - Edible coatings on fresh produce, meats, and cheese.<br>- Temporary antimicrobial surface layers on food-contact materials. | [38] |
| **Spray coating** | Plastic/metal sheet or food surface + chitosan solution (often low viscosity) | - Select proper spray nozzle, pressure, and spray time.<br>- Control droplet size and spraying distance.<br>- Maintain temperature (≈20–30 °C) and relative humidity (≈40–60 %) for uniform drying. | - Simple continuous procedure suitable for conveyor systems.<br>- Fine control of coating thickness and area.<br>- Can treat large surfaces quickly. | - Coatings often have low mechanical strength.<br>- Barrier properties depend strongly on process control.<br>- Overspray and material loss increase cost. | - Sprayable edible coatings for fruits and vegetables.<br>- Thin antimicrobial layers on packaging films or trays. | [38] |
| **Layer-by-layer (LbL) assembly** | Solid substrate (glass, plastic, metal) + alternating cationic (like chitosan) and anionic polymers or particles | - Prepare dilute polyelectrolyte solutions at moderate pH (typically 4–6).<br>- Immerse substrate alternately in oppositely charged solutions with rinsing between steps.<br>- Number of bilayers controls thickness and functionality. | - Precise nanometer-scale control over thickness.<br>- Ability to incorporate a wide variety of functional materials (enzymes, nanoparticles, proteins).<br>- Excellent mechanical strength and barrier properties when many bilayers are used. | - Requires specific equipment and multiple immersion steps.<br>- Time-consuming and less suited for high-throughput industrial scale.<br>- Process sensitive to pH, ionic strength, and solution stability. | - Multilayer active coatings on films, implants, or sensors.<br>- Tailored release systems for antimicrobials or antioxidants. | [39] |
| **Solution blending (film casting of blends)** | Chitosan + other polymers (like starch, PVA, proteins) + plasticizers/other additives in common solvent | - Select solvent compatible with all components (often aqueous acetic acid).<br>- Maintain predetermined temperature and humidity.<br>- Mix until homogeneous, then cast and dry like solvent casting. | - Good barrier properties and lower oxygen permeability when properly formulated.<br>- Improved mechanical strength and flexibility relative to neat chitosan.<br>- Enhanced compatibility and antibacterial properties via synergistic effects. | - Film properties strongly affected by solvent quality and drying conditions.<br>- Possible phase separation or incompatibility between polymers.<br>- Residual solvent may remain if drying is incomplete. | - Composite films for food packaging with tuned strength and barrier (like chitosan–starch, chitosan–PVA). | [41] |
| **Extrusion** | Chitosan (often thermoplastic or blended, like with starch or PVA) + plasticizers/additives | - Use solvent free or low-solvent formulations compatible with screw extruders.<br>- Control barrel temperature profile, screw speed, and die temperature.<br>- Maintain external conditions to prevent degradation. | - Lower cost per unit and suitable for large-scale production.<br>- Continuous process compatible with industrial film lines.<br>- Can improve oxygen barrier and mechanical performance with proper additives. | - Requires careful optimization; chitosan alone is difficult to extrude.<br>- Properties highly sensitive to processing parameters.<br>- May contain residual additives or plasticizers and be more sensitive to humidity. | - Industrial-scale biodegradable films and blends with starch or other polymers for packaging. | [43] |
| **Ionic gelation** | Chitosan solution + multivalent polyanion (like TPP, citrate, alginate) | - Control chitosan and polyanion concentration, mass ratio, pH, temperature, and stirring rate.<br>- Dropping or spraying polyanion into chitosan (or | - Mild, water-based process without organic solvents.<br>- Enables formation of nanoparticles, microgels, or thin films with improved mechanical stability. | - Sensitive to small variations in pH and mixing conditions, leading to size or thickness variability.<br>- Crosslinking can reduce film flexibility. | - Fabrication of nanoparticles for active coatings or as packaging polymers of films.<br>- Controlled-release | [45] |
| | | vice versa) to form instantaneous gel networks. | - Good encapsulation capability for bioactives. | | systems in food or biomedical applications. | |
| **Emulsion droplet coalescence** | Emulsifiers + polymer (chitosan) + proteins/particles; oil or water phase depending on formulation | - Control viscosity, pH, ionic strength, temperature, and droplet size.<br>- Coalescence or crosslinking of droplets forms particles or continuous films upon drying. | - Good for encapsulating hydrophobic activities (like essential oils).<br>- Allows formation of porous films or particles with high surface area. | - Process control is critical; small changes in viscosity or ionic strength strongly affect results.<br>- Can be more complex than direct casting. | - Delivery of lipophilic antimicrobials or antioxidants in chitosan matrices.<br>- Active and functional coatings or films. | [58] |
| **Electrospinning** | Chitosan + PVA or other carrier polymer + dilute acid (like 2 % acetic acid | - Adjust chitosan/PVA concentration to control solution viscosity.<br>- Apply high voltage between needle and collector.<br>- Set capillary-to-collector distance (≈10–15 cm) and flow rate to obtain uniform nanofibers. | - Produces nanofibrous mats with very high surface area.<br>- Improved oxygen barrier, rapid drying, and tunable porosity.<br>- Good for loading bioactives and creating multilayer structures. | - Requires high-voltage equipment and careful safety controls.<br>- Lower throughput than casting or extrusion.<br>- Pure chitosan is hard to electrospin; it usually needs a carrier polymer (like PVA). | - Active nanofibrous coatings and multilayer films for high-value foods.<br>- Scaffolds for biomedical and regenerative applications. | [48, 50, 51] |

### 2.3. Additives and Formulation Design Considerations

While chitosan has natural antibacterial activity, film-forming capabilities, and biodegradability, it rarely meets all the functional requirements for food packaging applications in its unmodified form [59]. As a result, additives are widely used in synthesis or postprocessing to control mechanical strength, barrier performance, stability, and functional activity. The timing, selection, and incorporation method of the additives are determined by the physicochemical restrictions of chitosan, the processing conditions, and the desired functional role of the resulting material. Understanding these factors is critical for the logical design of chitosan-based packaging systems [11, 60].

#### 2.3.1 Influence of Chitosan Structure on Additive Incorporation

The chemical structure of chitosan significantly influences additive compatibility and incorporation processes. Chitosan is a linear polysaccharide consisting of β-(1→4)-linked D-glucosamine and N-acetyl-D-glucosamine units. It features primary amine and hydroxyl functional groups that facilitate hydrogen bonding, electrostatic interactions, and covalent modification [60]. Functional groups act as primary interaction sites for additives, including plasticizers, phenolic compounds, nanoparticles, and cross-linkers [11, 61].

Two inherent parameters, namely, degree of deacetylation (DD) and molecular weight (MW) determine when additives can be inserted. A greater DD increases the density of protonated amine groups under acidic circumstances, which increases cationic charge density under acidic pH, strengthening electrostatic complexation with anionic additives (like alginate, pectin, TPP) and enhancing adsorption of some polyphenols via combined electrostatic and hydrogen-bonding interactions [62]. Higher MW chitosan solutions, on the other hand, have greater viscosity, which might impede additive dispersion and cause aggregation if additives are applied too late in the process. As a result, additives designed for bulk modification are frequently administered during the first dissolving stage, when molecular mobility is greatest [30].

Chitosan's poor solubility further limits additive timing because chitosan is hydrophilic, and additives that reduce swelling (hydrophobes, crosslinkers, nanofillers) are often prioritized when water-vapor barrier is critical. The polymer is insoluble in neutral water and only soluble in mildly acidic aqueous solutions like acetic acid. Additives that are unstable in acidic conditions or weakly soluble in aqueous systems cannot be safely supplied during solution preparation, thus they are inserted after film formation along surface coating or sequential deposition procedures [63].

#### 2.3.2 Processing Method as a Determinant of Additive Timing

The synthesis process utilized to produce chitosan films or coatings has a direct impact on when additives can be applied. Each processing method imposes unique chemical, thermal, and mechanical restrictions that influence additive stability and distribution [62]. To facilitate homogeneous dispersion, solvent casting additives are often incorporated directly into the chitosan solution prior to film production [62]. This method is ideal for plasticizers, antioxidants, and antibacterial compounds that remain stable during solvent evaporation and drying. However, prolonged drying times and acidic conditions might cause sensitive additives to degrade or volatilize, necessitating alternative integration procedures [64, 65].

Similarly, solution blending involves the codissolution of chitosan and additives in the same solvent. This approach promotes strong intermolecular interactions and homogenous bulk modification, but it necessitates careful control of solvent quality and processing conditions to avoid phase separation or residual solvent retention [66]. If applied prematurely, additives that significantly modify hydrogen bonding networks may have an effect on film morphology.

In contrast, layer-by-layer (LbL) assembly permits additives to be inserted after the initial chitosan layer has formed. This sequential deposition method allows for precise spatial control of additives while limiting exposure to harsh processing conditions [67]. As a result, LbL techniques are ideal for functional additives like antimicrobials and sensing chemicals that must maintain their activity [68].

To guarantee equal mixing in extrusion-based processing, additives are usually added before thermal processing. Because extrusion exposes materials to heat and shear, additives should be thermally stable, shear-stable, and compatible with the chitosan blend. Heat-sensitive bioactive chemicals are frequently omitted from early extrusion phases, instead being administered postextrusion via surface coating or subsequent treatments [65].

### 2.3.3 Functional Role of Additives and the Incorporation Strategy

The functional role of an additive has a major influence on its incorporation stage. Plasticizers, antibacterial agents, antioxidants, nanofillers, and cross-linking agents all have distinct interactions with the chitosan matrix and respond differentially to processing conditions [69]. Plasticizers, such as glycerol and sorbitol, are commonly used during solution preparation because they disrupt intermolecular hydrogen bonding between chitosan chains, increasing flexibility and elongation [70]. Early inclusion ensures consistent alteration of mechanical characteristics. However, plasticizer addition may increase water sorption and water-vapor transmission and can introduce migration concerns, requiring optimization of loading and compatibility. The performance of antimicrobial and antioxidant additives is highly processing dependent. Some phenolic chemicals and essential oils can be added directly to chitosan solutions, whereas others are more likely to oxidize, volatilize, or degrade after drying. In these circumstances, postformation surface integration or encapsulation into the polymer matrix is frequently used to preserve functional efficacy [71, 72].

Recently conducted studies have underscored the significance of additive incorporation processes in chitosan-based active packaging. For instance, in their studies, Fan et al. [219] in 2025 examined zein-chitosan Janus films loaded with emulsified tannic acid and cinnamon essential oil particles, aimed at the control of the release of the active compounds. In their works, they showed the influence of the incorporation methods used in preparing chitosan films on the morphology and properties of the films. Wang et al. [220] in their works in 2026 used chitosan films with the incorporated zein-inulin stabilized thyme essential oil emulsions. The study by these researchers also provided evidence that incorporation methods used influence the morphology, thermal stability, barrier properties, and antioxidant activity of the films. The research highlights the importance of additives determining the properties of the packaging films. However, their characteristics in terms of food safety and sustainability of the introduction of the additives still require evaluation.

Nanoparticle-based additives create extra dispersion and aggregation issues. These additives are typically supplied at early solution stages with controlled agitation to promote uniform dispersion, however postformation deposition methods may be used when aggregation hazards are substantial [73]. Cross-linking agents are usually added during the posttreatment step to prevent premature gelation or excessive viscosity rises that might interfere with processing. Controlled crosslinking stabilizes the polymer network while maintaining processability [74]. Recent encapsulation and emulsion-based approaches further demonstrate that the timing and spatial distribution of additives can influence their stability, interactions with the chitosan matrix, and release behaviour during packaging use.

Additives added to chitosan to improve its physical and barrier characteristics include either polymer–polymer blends, nanoparticles, or posttreatment of coatings. Filler is very thin and usually functions to enhance the quality and capability of chitosan; it can increase the tensile modulus and gas transport tortuosity through the use of different mechanisms of chemical bond formation. Nanofillers can improve gas-barrier and mechanical performance by increasing diffusion tortuosity, reinforcing the polymer network, and modifying interfacial interactions. Their effectiveness depends on dispersion quality, interfacial compatibility, loading level, and stability during storage and disposal; covalent bonding is not always required. Because of these limitations, considerable time has been spent examining combinations of formulations of chitosan and poly(vinyl alcohol), which utilize both silica-based fillers (or cross-linking agents) or chitosan/poly(vinyl alcohol), with varying metals being evaluated [75]. The performance of hybrid formulations is strongly affected by moisture conditions and/or additives and by their method of preparation [75, 76]. Several different types of common inorganic and organic additives to chitosan are discussed in detail in **Table 3**.

Relative to conventional petrochemical plastics, chitosan offers a distinct combination of waste-derived feedstock potential, inherent antimicrobial functionality, and film-forming capability under mild acidic conditions. However, these advantages are counterbalanced by moisture sensitivity, limited mechanical robustness in neat films, and a strong dependence on degree of deacetylation, molecular weight, solvent environment, and additive selection to achieve application-relevant performance. In practice, the same molecular and formulation variables that improve tensile strength, barrier properties, or active functionality can also increase processing intensity, chemical input, or end-of-life uncertainty [76]. These coupled material–process relationships motivate the life-cycle interpretation developed in the following sections.

**Table 3.** Common additives used to enhance the properties of chitosan films for food packaging.

| Additive | Process(es) of Production of Chitin with Additive | Properties of composite film in comparison to chitosan | Importance in Chitosan Nanomaterials for Food Packaging | Advantages | Disadvantages | References |
|---|---|---|---|---|---|---|
| **Metal/Metal Oxide Nanoparticles ($TiO_2$, ZnO, and Ag)** | - Casting and in situ synthesis for $TiO_2$.<br>- Evaporation–condensation, chemical reduction, or biogenic green synthesis for AgNPs and ZnO NPs.<br>- Ultrasonication and dispersion under mild heat for uniform distribution. | - Strong antimicrobial and antifungal activity against *E. coli*, *S. aureus*, *Listeria monocytogenes*.<br>- Improved UV-light barrier and oxygen barrier properties.<br>- Increased shelf life and oxidation resistance. | Enhances active food packaging performance by extending food shelf life through antimicrobial and UV-blocking activity. Improves product safety and reduces spoilage. | - High antimicrobial efficiency.<br>- Can provide effective antimicrobial functionality at low loadings; biocompatibility and environmental effects are formulation-, concentration-, and exposure-dependent.<br>- Photocatalytic degradation resistance.<br>- Retains mechanical integrity and transparency at optimal nanoparticle load. | - Nanoparticle aggregation reduces uniformity.<br>- Metal ion leaching can raise safety concerns.<br>- Energy-intensive synthesis, potential cytotoxicity at higher concentrations. | [77, 78] |
| **Nanocellulose** | - Carboxylation or TEMPO oxidation of cellulose fibers followed by shear-mixing into chitosan dispersions.<br>- Vacuum filtration and solvent casting. | - Increased tensile strength, Young's modulus, and stiffness.<br>- Enhanced optical transparency (transmittance >85%).<br>- Lower water vapor permeability. | Provides mechanical reinforcement and transparency while maintaining biodegradability, making it a sustainable substitute for petroleum plastics. | - Lightweight and renewable.<br>- Reinforces mechanical and thermal properties.<br>- Enhances barriers to gases and oils.<br>- Fully biodegradable. | - Longer preparation and dispersion uniformity issues.<br>- Reduced ductility due to rigid fibril structure. | [79, 80] |
| **Essential** | - Disc diffusion and | - Increased antimicrobial and | Can provide antimicrobial | - Natural and organic. | - Property variability | [71, |
| **Oils/Extracts (e.g., oregano, thyme, cinnamon, and rosemary essential oils)** | solvent casting methods with micro/nanoemulsion incorporation.<br>- Direct blending or encapsulation of essential oils before film formation. | antioxidant activities.<br>- Reduced water vapor transmission rate (WVTR).<br>- Sometimes decreased tensile strength depending on the oil's composition.<br>- Potentially enhanced antimicrobial and antioxidant activity, depending on oil composition and loading.<br>- Water-vapor barrier properties may improve or deteriorate depending on formulation and dispersion.<br>- Tensile strength and elongation may increase or decrease with the essential oil and incorporation method. | and antioxidant functionality in active packaging, with potential food-preservation benefits depending on active-compound release, formulation stability, and food-storage conditions. | - Enhances aroma quality and antimicrobial efficacy.<br>- Plant-derived active compounds; food-contact safety and environmental behaviour depend on composition, concentration, and exposure conditions. | depending on oil type.<br>- Volatile and oxidation-sensitive can reduce storage stability.<br>- Often reduces tensile strength. | 81] |
| **Polyol Plasticizers (glycerol, sorbitol, etc.)** | - Solution casting or extrusion blending.<br>- Added to chitosan film-forming solution before drying. | - Increased flexibility, elongation at break, and film transparency.<br>- Reduced brittleness and improved processability.<br>- May reduce tensile strength and gas barrier at higher loadings. | Used to improve flexibility and film-handling in packaging applications without sacrificing biodegradability. | - Derived from renewable sources.<br>- Enhanced elasticity and formability.<br>- Improved barrier to $O_2$ at optimal concentration. | - Increased hydrophilicity leads to higher water permeability.<br>- Plasticization reduces tensile strength. | [82-84] |
| **Clay** | - Layer-by-layer assembly or direct dispersion in solution-cast chitosan matrix (1–5 wt%). | - Increased tensile strength and modulus up to an optimal loading.<br>- Decreased water vapor permeability (WVP).<br>- Enhanced antioxidant activity and light barrier. | Serves as a low-cost, nontoxic reinforcement improving strength, thermal stability, and barrier functions, especially for oxygen and moisture. | - Abundant and inexpensive.<br>- Nontoxic and biocompatible.<br>- Improves mechanical and barrier properties. | - Overloading causes opacity and brittleness.<br>- May affect food sensory quality if migration occurs. | [85] |
| **Polyvinyl Alcohol (PVA)** | - Solution casting or electrospinning of chitosan/PVA blend at varying weight ratios. | - Improved flexibility, tensile strength, and elongation at break.<br>- Reduced oxygen permeability and good transparency.<br>- Increased water absorption. | Produces flexible, strong, and transparent films suitable for oxygen-sensitive food packaging. | - Excellent oxygen barrier.<br>- Biocompatible and biodegradable.<br>- Good thermoplastic and blending properties. | - Susceptible to humidity.<br>- Increased water permeability limits wet-food applications. | [75, 86] |
| **Silica-based ($SiO_2$)** | - Sol–gel synthesis or direct dispersion with casting methods. | - Increased tensile strength and stiffness.<br>- Reduced WVP and increased moisture resistance.<br>- Enhanced UV and thermal stability. | An important moisture barrier additive prevents condensation and microbial growth, extending product freshness and visual clarity. | - Environmentally benign.<br>- Thermally stable.<br>- Biocompatible and antimicrobial. | - Residual silica particles after degradation need disposal assessment. | [75, 87] |
| **Proteins (gelatin, quinoa, whey, soy)** | - Blend by casting method using acidic aqueous solutions. | - Increased elongation at break, improved film flexibility.<br>- Enhanced antioxidant and emulsification properties.<br>- Decreased tensile strength but edible and transparent. | Enables development of edible, nontoxic packaging films for temporary protective coatings or inner wrappers. | - Edible, sustainable, and renewable.<br>- High antioxidant and antimicrobial action.<br>- Biocompatible and biodegradable. | - Poor water resistance.<br>- Allergenicity potential in some proteins.<br>- Reduced mechanical strength. | [22, 88] |
| **Flavonoids** | - Added a chitosan solution before solvent casting or electrospinning for active films. | - Enhanced antimicrobial and antioxidant activity.<br>- Reduced WVP and oxygen permeability.<br>- Improved color stability and surface heterogeneity. | Serves as a natural antioxidant and antibacterial agent improving food quality retention by preventing lipid oxidation and microbial spoilage. | - Prolongs shelf life.<br>- Nontoxic and biocompatible.<br>- Adds antioxidant activity and UV protection. | - Relatively expensive compounds.<br>- Light- and heat-sensitive.<br>- Possible uneven distribution without surfactants. | [89-91] |

## 3. Life cycle interpretation through material and process decisions

The production process and environmental impact of chitosan-based food packaging is composed of three intertwined elements: the requirements for upstream production processes, the attributes and functionalities of the packaging itself, and the post-use disposal management processes. The extraction process of chitosan usually needs to go through a lot of chemical and thermal operations, which often consumes a lot of water and energy. When the formulation and manufacturing processes are considered, the environmental burdens become even more drastic because of the raw materials required for formulation processes. The assessment of such impacts cannot be carried out in isolation. For example, a particular formulation may imply more manufacturing impacts than others, but may also be able to provide a substantial good environmental impact through reducing the spoilage of the expensive product significantly.On the other hand, a readily degradable film can be detrimental to the environment if it calls for much more raw material and energy to deliver the same level of protection, or if it leads to greater food wastage. [92-94] For this reason, the present review evaluates chitosan-based packaging in terms of three interrelated dimensions, namely: (i) inventories of materials and process, (ii) food protection efficiency, along with system-level implications, and (iii) safety, as well as end-of-life outcomes, specific for each formulation. The evidence collected is studied with particular reference to the characteristics of functional units as well as boundaries of the system of study, as well as the level of uncertainty associated with the quality of inventory data.

Materials made from chitosan have lower ecological footprints than many conventional plastics derived from fossil fuels do, provided that they are produced from reused waste biomass and discarded appropriately after use. Calculating the comprehensive sustainability of chitosan throughout its entire life cycle instead of just considering its biodegradability in a landfill requires conducting a life cycle assessment (LCA) on the products made from a material, the processes that create them, their end use, and their end-of-life disposal/recycling methods. Variables that influence the environmental profile of chitosan-based food packaging include, but are not limited to, the chemicals used in manufacturing, the energy expended to produce the packaging films, the water consumed during production, and their required performance attributes when finished [14, 93].

When the environmental impact of chitosan packaging materials (also known as chitosan polymer products) is assessed, the transportation of materials, including the extraction and processing of shell waste, the extraction and processing of reagents used in chitosan production, and the transport of chitosan packaging to retail locations, can contribute greatly to the overall life-cycle burden of the product. Unfortunately, this information is often not readily available in publications, and much of the published research tends to focus primarily on extraction and processing rather than logistics associated with transporting materials. This results in many assessments having a considerable underrepresentation of the burdens associated with transporting materials and finished products associated with chitosan polymer products [65, 94].

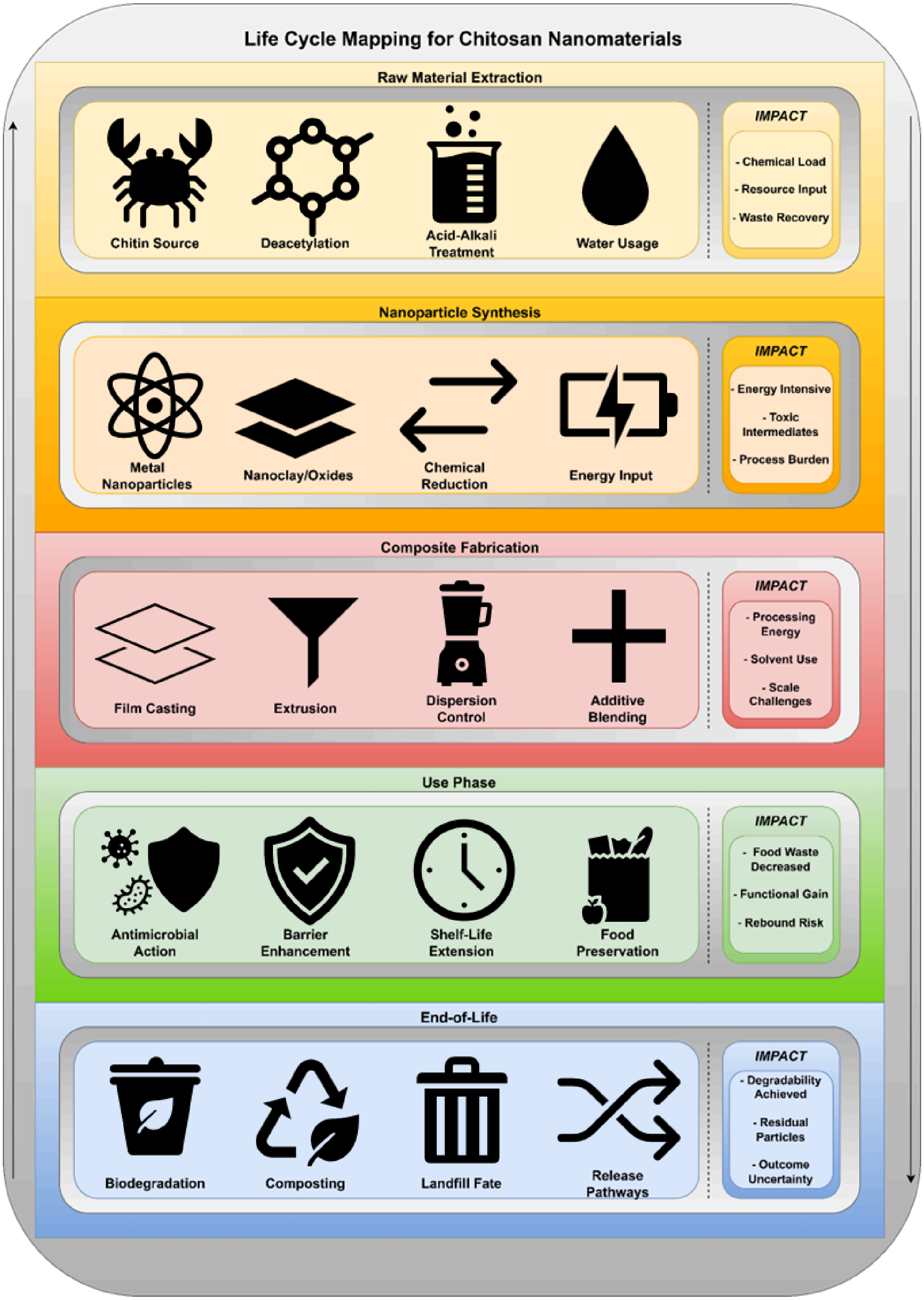

**Figure 3. Life cycle mapping of chitosan-based nanomaterials for food packaging applications.** *The figure outlines the sequential stages from raw material extraction to*

*end-of-life, highlighting key processing steps and associated environmental impact signals at each stage. Raw material extraction involves chitin sourcing and chemical processing, characterized by resource inputs and chemical load. Nanoparticle synthesis introduces additional energy and chemical demands, contributing significantly to process intensity. Composite fabrication encompasses film formation and additive integration, which are associated with solvent use and scalability constraints. During the use phase, functional properties such as antimicrobial activity and barrier enhancement contribute to food preservation and potential reduction in food waste. End-of-life pathways include biodegradation and disposal scenarios, with outcomes influenced by degradation behavior and environmental conditions. Collectively, the data in the figure emphasize the uneven distribution of environmental burdens and functional benefits across life cycle stages.*

### *3.1. Raw material origin: waste valorization* **vs** *dependency creation*

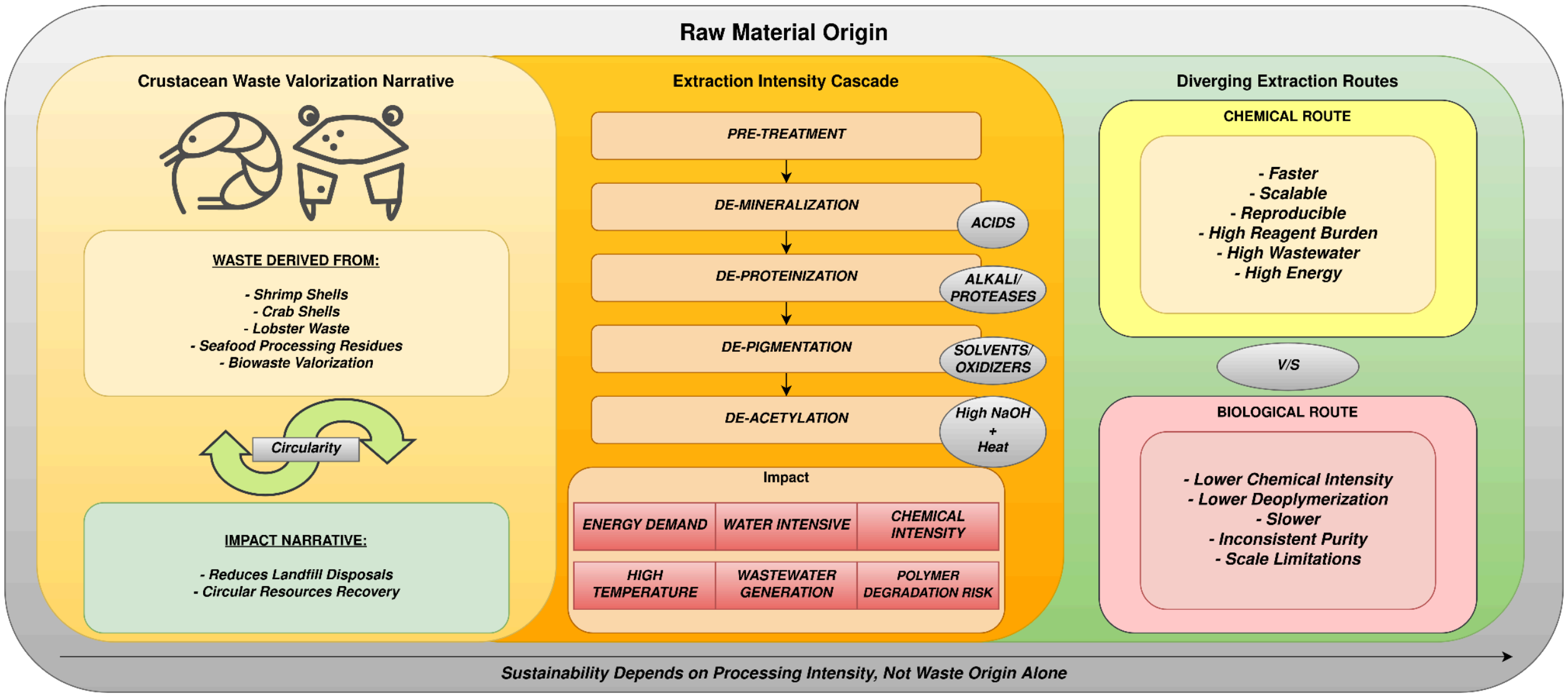

**Figure 4. Conceptual framework illustrating upstream raw material extraction pathways and associated sustainability trade-offs in chitosan production from crustacean waste streams.** *The figure highlights the sequential extraction intensity cascade, including pretreatment, demineralization, deproteinization, depigmentation, and deacetylation, alongside their associated chemical, water, and energy burdens. Parallel comparison between chemical and biological extraction routes demonstrates the balance between scalability, process efficiency, and environmental intensity. While crustacean shell valorization supports circular resource recovery and landfill diversion, the framework emphasizes that sustainability outcomes remain strongly dependent on upstream processing intensity rather than waste-derived origin alone.*

Before chitosan can be produced, chitin must first be isolated from shell biomass and then deacetylated to convert part of the N-acetyl-D-glucosamine units into D-glucosamine units (Figure 4). Although chitin can be extracted from fungi and squid, it is most commonly derived from crustacean shells because of its high chitin content and widespread availability [95]. There is a variation in the quantity and composition of processing residues generated from crustaceans. There are different reported chitin fractions in crustacean shell materials, which have a representative range of 15-40%. Individual feedstocks may have values higher or lower than the range because of shell composition and type of analysis [216]. However, it is critical to note that the valorization of this waste could result in a dependency on resources that could evolve exploitatively. Each step of raw material extraction requires energy and resources that can be environmentally and economically costly and in terms of the quality of the end chitosan product itself. Trade-offs across extraction routes and downstream impacts on chitosan properties are discussed in Section 4.

To obtain from crustacean waste to chitin, both chemical and biological processes can be used. Industrial production is predominantly chemical, owing to shorter processing times and established supply chains, despite higher chemical demand and effluent burdens. The entire process from waste to chitin involves five general steps (Figure 4):

- *Pretreatment*
- *Demineralization*
- *Deproteinization*
- *Decolorization/depigmentation*
- *Deacetylation*

Shells are typically washed to remove adhering tissue and salts and then size-reduced to increase surface area for subsequent treatments. The next step is demineralization to remove minerals, especially $CaCO_3$ ; either sequence is reported in the literature; however, the process order can influence the kinetics, reagent consumption, and partial depolymerization depending on the conditions. The most common method to achieve this is through acid treatment using chemicals such as HCl, HCOOH, $HNO_3$, $CH_3COOH$, and $H_2SO_4$. This process requires extraction temperatures of about 100 °C or above for longer periods of time. At the molecular level, acids react with minerals to produce water, carbon dioxide, and soluble salts in the presence of acid. The solid chitin was filtered out from the acid and salt solution and then was washed with deionized water. Acid converts $CaCO_3$ into soluble salts, releasing $CO_2$ and enabling the separation of a chitin-rich solid. Mineral byproducts can be used industrially for some pharmaceuticals, agricultural products, and paper products [96]. However, many of the product derivatives of chitin require further processing [96, 97].

Chitin extracted from the waste products produced by crustaceans is performed using either chemical or biological processes; however, the chemical method remains the predominant method of industrial production because of its faster processing time and ability to use existing processing equipment, despite having greater quantities of reagents and requiring more energy to perform the extraction process. Recovering purified chitin from shell waste typically requires four stages: pretreatment of shell waste, demineralization of the shell matrix, removal of proteins using a solubilizing agent, and removal of pigments from chitin using bleaching. Typically, the first step in recovering a purified form of chitin from crustacean waste is to wash the shells to remove residual tissue prior to reducing their size so that efficient mass transfer can take place during the extraction process. The sequence and intensity of each of these processes significantly affect the amount of reagent consumed in the extraction process, the overall recovery of the chitin polymer, and its degradation during extraction [98].

Demineralizing the shell is primarily a method of removing calcium carbonate and other minerals from the shell matrix. Most commonly, mineral removal occurs through the use of acid (hydrochloric acid or organic acids) during the demineralization step, which converts the calcium carbonate in shells into $CaCO_3$ into soluble calcium salts, $CO_2$, and water, liberating $CO_2$ and resulting in a chitin-rich solid remaining after calcium carbonate has been removed from the shell matrix. Although there are environmental benefits associated with removing minerals from shells to facilitate further purification processes, the benefits associated with the use of acid to remove minerals must be offset against the operational cost of using acids and the amount of wastewater generated through acid washing and neutralizing the acidic effluents generated during the washing of crustacean shells [98]. Side streams generated from the demineralization process that contain high levels of minerals may provide opportunities for valorization, such as use in agricultural or paper applications, but may require further purification or processing prior to reuse, thereby decreasing their overall environmental advantages compared with other valorization options available for crustacean shell waste [99]. For hydrochloric-acid demineralization, the ideal stoichiometric reaction is $CaCO_3 + 2HCl \rightarrow CaCl_2 + H_2O + CO_2$, corresponding to approximately 0.73 kg of pure HCl per kg of $CaCO_3$ removed. Actual reagent consumption is higher and depends on shell mineral content, acid concentration and excess, mass-transfer efficiency, washing, and neutralization requirements. This theoretical requirement should therefore be distinguished from industrial inventory data for the complete chitosan-production process.

Protein removal reduces allergenic residues and improves polymer purity and performance consistency, enabling safer use in biomedical and food-packaging applications. Deproteinization is typically achieved through alkaline treatment, which hydrolyzes protein–matrix interactions and solubilizes proteins; however, harsh conditions may reduce molecular weight. The resulting protein-rich liquors may be explored for valorization (e.g., as animal feed or fertilizer), although high salinity and residual processing chemicals can limit their applicability. Sodium hydroxide (NaOH) is the most commonly used base industrially, although a wide range of alternative alkaline solutions may also be used [100].

Food packaging applications may require the use of depigmentation if one wishes to use light-colored materials, especially if the factors of appearance, transparency or flexibility of the formulation are relevant. Many common reagents used in this process include solvents such as ethanol and acetone, in addition to various oxidizing agents, e.g., sodium hypochlorite (NaOCl), hydrogen peroxide ($H_2O_2$) and potassium permanganate ($KMnO_4$). However, the use of depigmentation is not environmentally neutral, as the increase in chemical demands through the use of additional solvents or oxidizing agents creates additional washing and waste treatment requirements. The high temperatures involved in aggressive bleaching processes can also change the properties of the polymer used for food packaging. Therefore, depigmentation should be considered only when necessary for the intended application rather than as an automatic requirement for all manufactured products [101].

Chitin can be converted into chitosan via a deacetylation process using concentrated sodium hydroxide (NaOH) at high temperatures. The factors affecting the degree of deacetylation (DD) and molecular weight of chitosan are the concentration of NaOH, reaction time, and temperature. These factors strongly influence the pliability, film-forming ability, mechanical ability, and barrier functionality. Severe deacetylation of chitosan requires more reagents and energy. Additionally, the extent of chain scission (the splitting of long molecular chains) increases with severe conditions. As a result, deacetylation must balance the functional properties of the final chitosan form with the environmentally desirable attributes of chitosan produced using moderate versus severe processing conditions. While chemical processes dominate the current industry for chitin extraction and its conversion to chitosan, various biological processes can be employed either independently or in combination with chemical methods to produce chitosan. Much research suggests that biological methods have far less of a negative effect on the environment, and adoption at scale remains limited, largely because of longer processing times, variability, and downstream purification requirements [100, 101].

One commonly used method is enzymatic deproteinization. Purified or crude proteases, which perform more efficiently at a lower cost, can most often be derived from bacteria and fish viscera, but their overall results are not as thorough as those of chemical methods. Approximately 5–10% of residual protein is left behind after biological enzyme treatment, so studies advise a small NaOH treatment to complete deproteinization. In this case, the environmental impact and chitin structure depolymerization would be reduced by the use of the enzyme process, but there would be more steps to apply, which can become costly. Complete purity from protein is not necessary for food packaging applications, but lower residual protein is associated with higher-quality chitin. Additionally, research has shown that demineralization before enzymatic deproteinization can increase the efficiency of the latter process. This is because if minerals are present in the chitin matrix, they can prevent proteases from accessing some proteins [102, 103].

Microorganism fermentation methods can also supplement chemical and enzymatic processes as a generally less expensive option. Selected strains of bacteria that produce lactic acid (LA) and proteases such as *Lactobacillus* sp. are commonly used for both the demineralization and deproteinization of chitin. The LA produced reacts with calcium carbonate, the primary mineral in chitin, and forms removable calcium lactate, while extracellular proteases aid in deproteinization. In addition, a low pH is created during the production of LA, which prevents the growth of spoilage microorganisms. It is common for other bacterial strains that produce more proteases to be added to allow for deproteinization as well. However, many factors can affect the end product of chitin, such as the initial pH, temperature, glucose input, and incubation time [100, 104-107].

Fermentation can be used as an alternative or supplement to chemical methods. Extraction through fermentation involves the use of specific strains of microorganisms (such as Lactobacillus) to produce lactic acid (and in certain cases, have proteolytic enzymes), allowing for a portion of the raw material(s) to be demineralized in one step. When Lactobacillus produces lactic acid, it binds to calcium carbonate to form calcium lactate, and extracellular proteases can assist with the removal of proteins. In addition, the lactic acid produced by microbes creates an acidic environment, which inhibits the growth of spoilage microorganisms during fermentation. Since fermentation has the potential to reduce the reliance on strong acids/alkalis for sustainability reasons, the success of the fermentation process depends on several key process parameters, including temperature, pH, glucose level, strain, and incubation time. Thus, this very highly sensitive range of process parameters significantly affects the consistency of purification and the ultimate properties of the harvested chitin [106-108].

It is also possible to convert chitin into chitosan via enzymes called chitin deacetylases, which are obtained from various fungal and insect sources. Enzymatic methods to produce chitosan include direct hydrolysis of the N-acetamido bond to yield chitosan without undergoing a concentrated alkali process, which is a milder method. Many factors influence variables such as the enzyme source, operating parameters (pH), and level of deacetylation in any type of enzymatic method. Although there is currently no commercial production or recovery of chitin deacetylases, all the data available at this time come from laboratory sample production [109, 110]. Physical extraction processes, such as ultrasonication and microwaving, have been used to increase reaction rate time and decrease the processing time or reduce the amount of chemicals used for process-excited extraction. These physical extraction processes have the potential to increase extraction efficiency and improve quality, although their overall sustainability depends on energy consumption for each of the processes and their scalability. Currently, there are no widely used commercial applications for these physical extraction processes [100, 111].

Chitin and chitosan can both be acquired from crustacean shell waste through different methods of extraction. Each of these methods of extraction has a different degree of environmental burden, and the quality of the final product also depends on the number of additional steps that are required to convert chitin to chitosan. Therefore, one cannot determine the sustainability of chitosan purely because it comes from a waste source and must consider the degree of chemical or biological processing that occurs to achieve the required purity and functional characteristics of the material. For this reason, raw material valorization and process design are inseparable when evaluating the life cycle impacts of chitosan and related products [15, 112].

### *3.2. Manufacturing pathway choices: hazard reduction* vs *chemical burden*

Food-packaging-grade chitosan materials have been manufactured under conditions designed to meet the requirements for compositional purity, functional capabilities, and regulatory safety. However, meeting these requirements may lead to increased consumption of reagents, increased demand for heating energy, and increased intensity of wastewater treatment. These variables create a trade-off between reducing hazards from packaging and their overall environmental impact. For example, optimizing package tensile strength, elongation and barriers may require cross-linking, the use of additives, or more extensive posttreatment methods. All three of these options yield improved functional performance of the final package but decrease biodegradability or increase the cost of the final package. Therefore, many manufacturing decisions can impact not only the performance of the resulting product but also the establishment of environmental hotspots for the entire packaging system [113].

Some manufacturing procedures help reduce the risk of biological and allergenic hazards, as outlined in Section 3.1. For example, deproteinization, depigmentation, and sterilization can contribute to reducing risk; however, they do not eliminate all the risk associated with the production process. Instead, these procedures may just shift the risk to other areas of production, such as chemical handling, large quantities of salt-rich effluent, or high energy consumption for neutralization. Thus, when determining the safest or least hazardous way to produce chitosan and chitosan-based films, chemical demands, energy consumption, and a comprehensive risk analysis based on multiobjective models should be completed on an interrelated basis as opposed to using a simple yes/no decision matrix. Understanding this is crucial to understanding why some alternatives that are considered 'greener' have been restricted because they cannot be produced on a larger scale because of product quality inconsistency and regulatory requirements for quality assurance [114].

Following the purification process, the chitosan films are produced using more acidic and alkaline solvents. During film formation, a viscous solution is formed by mixing purified chitosan in diluted acid. For example, pure acetic acid was combined with deionized water to create a 1.0% (w/v) acid solution. The chitosan powder was dispersed in the aqueous acetic acid solution at the required concentrations to prepare the film-forming solutions. Generally, the solution is then poured over a flat surface, and the solvent is allowed to dry. The dried film then undergoes neutralization to remove any residual acid by immersing the film in an alkaline solution or purified water and drying once again [115]. Moreover, residual acetic acid remaining after solvent-based chitosan film formation should be considered when evaluating food-contact applications because incomplete removal may affect film properties, sensory quality, and potentially migrate into the packaged food [115, 221]. If these retained solvents are present, it creates concerns regarding health effects on humans and environmental impact. Some residual solvents, including acidic and alkaline trace impurities and even heavy metals, used within the manufacturing process make it difficult for the material to decompose or may contaminate the site of decomposition or downstream. These residual solvents could lead to increased risks of human health hazards and possible contamination. Depending on the type of residual solvent left behind, certain solvents can lead to direct toxicity if disposed of incorrectly [116].

One of the methods, as discussed in **Table 2** earlier, is solution or solvent casting, which involves dissolving chitosan in an aqueous solution of an organic acid; however, acetic acid is most commonly utilized. In industrial-scale methods, the volume of solvent used in casting increases, resulting in emissions and waste [117]. The most common method is acid–alkali extraction, which includes deacylation, demineralization, and deproteinization processes to produce thin chitosan films [117]. These processes use harsh acidic and alkaline solutions to dissolve chitosan into a solution and then deprotonate the amine groups. These processes prevent biological hazards by creating food-grade material; however, they involve high chemical loads and are water- and energy-intensive processes [118]. Acidic and alkaline solvents used on a large scale are readily available and inexpensive, which makes them excellent choices for mass production even when concerns are prevalent.

There are several caveats in the use of biowastes as well, especially for large-scale production; not only will it require more energy to produce a larger quantity, but manufacturers and industry owners may exploit the resource. This exploitation leads to overfishing, which impacts further steps in the material's life cycle and the future of chitosan film production. However, with the intent of mass supply, bioderived packaging may in turn aggravate carbon footprint release because of energy consumption and wastewater generation from the processes. [100]

The use of caution should also apply to the use of biowaste to develop pellet fuel (feedstock) for large-scale production. Crustacean shells are generally thought of as low-burden waste streams, but with the growing demand by industry for chitin created from crustacean shells, there may be a change in their economic value and how they are perceived as having an environmental burden-free status. The potential to generate sustainability through agro-waste valorization may be compromised by the increasing demand for energy, transportation and wastewater generated by increasing levels of production during this time. In addition, the assessment for the manufacturing expansion of bioderived packaging (to create sustainable packaging from waste) should not automatically be assumed to be of low impact; rather than this assumption, the evaluation should also consider the allocation or distribution of production, supply chain dependency during shipment and delivery of products to retail and consumer end-users, production process intensity and environmental impacts associated with production.

Energy demands in the production of chitosan packaging are an issue, even if the resulting chitosan material is considered "biobased". Many phases of the chitosan extraction process, such as deproteinization and deacetylation (upstream) and drying and posttreatment (downstream), may require high levels of heat input. Using fossil-fuel-based energy for these operations produces greenhouse gas (GHG) emissions on the basis of these processes. The extraction of chitosan and film formation produce concentrated amounts of hazardous wastewater that may require neutralization or disposal before it enters the environment. This contributes to both energy demand and environmental impact [119]. Therefore, if chitosan-based packaging is produced using nonrenewable energy and chemically intensive waste treatment systems, the sustainability benefits of chitosan decrease. From a life-cycle perspective, achieving food contact compliance and reducing the potential for hazardous waste disposal may generate increased carbon emissions, increased water use, and increased effluent treatment [120].

The mechanical, moisture, and barrier properties of chitosan materials have not been adequate to meet market demands for high-performance food packaging applications; therefore, multiple types of additives, such as plasticizers and reinforcers, have been utilized to enhance the performance properties of the chitosan matrix through methods such as cross-linking and the formation of nanocomposites during film processing. Additives can improve the mechanical strength and flexibility, as well as the oxygen or water vapor barrier properties of films, and in some cases, they can render the films antimicrobial to extend the shelf life of food products and improve the functionality of the packaging material. [118-120]

Although the additive-enhancement of chitosan will improve the performance characteristics, it can also have negative environmental and human health effects associated with the use of these types of additives. Depending on the quantity and type of additive employed, chitosan films made with additives may have a reduced degree of biodegradability and a changed end-of-life fate. Many biodegradation-resistant or poorly biodegradable additives remain in the environment after the chitosan matrix has degraded, including some metal-based nanoparticles used to enhance the barrier properties of films, thus posing a risk of environmental persistence, food safety to humans because of contact with packaging material, or the migration of the additive into packaged food. Consequently, while many of these additives significantly improve the practical performance of chitosan-based films, they may also generate new and significant sustainability risks throughout the life cycle of the chitosan-based film. A more detailed description of these types of trade-offs is provided in Section 3.3.

### 3.2.1 Quantifying hidden upstream burdens and carbon-footprint misperception

While chitosan is seen as a sustainable packaging option due to its place of origin and the fact that it biodegrades at the end of its life, there are a number of factors to consider when evaluating chitosan's sustainability claims which are overlooked, including upstream processes used to manufacture chitosan, such as: the method of processing chitosan from shellfish waste; and demineralisation, deproteinisation, deacetylation, washing, drying, and treating wastewater and solvents. Therefore, a product that is promoted as being environmentally friendly may, in reality, have a far greater environmental footprint than the projected sustainability benefit associated with that product has suggested [59, 121].

This illustrates that an intermediary processor accounts for a significant portion of the overall burden associated with a particular product or process. Muñoz et al. [18] conducted a consequential cradle-to-gate life-cycle assessment of industrial chitosan production in India and Europe using primary production data. The study considered the additional production of 1 kg of chitosan at the manufacturer's gate and used substitution to account for changes in competing feedstock uses and by-product management [18]. Their study showed that the main lifecycle environmental impact from the production of chitosan in India is derived from its parent material chitin and less from the waste shell. The life cycle impact of the production of chitosan in Europe has been attributed to coal based thermal and electrical energy in the more complicated international supply oriented to produce chitosan. Their study contradicts the generalised view that waste-derived origin provides documentary evidence of having a minimal climate burden. This study demonstrates the degree of variability that exists within a single product label [18]. For the supply chain tied to India, emissions from greenhouse gases from chitosan were reported to be approximately 12 kg $CO_2$-eq per kg of chitosan, while the supply chain connected to Europe had emissions of around 77 kg $CO_2$-eq per kg of chitosan (an increase of almost 6.5 times). These differences reflect variations in feedstock origin, processing configurations, energy supply, and product specifications. Importantly, the Indian and European-linked systems produced chitosan for different application markets; consequently, their reported environmental impacts should not be interpreted as a direct comparison of functionally equivalent products [18]. Furthermore, Muñoz et al. [18] concluded that the impacts of producing chitosan were greater than those of producing chitin, with chitosan contributing at least twice the greenhouse gas emissions attributable to production of chitin due to the need to deacetylate the chitin and for downstream processes to manufacture the final polymer product [18].

According to Muñoz et al. [18], the production inventory of India shows the need for resources required in conventional extraction and conversion. As per the inputs stated in the report, the total quantity of inputs in each of the processes is around 3.58 kg of active equivalent HCl, 7.00 kg of NaOH input, and 484 L of freshwater per kg of final chitosan, besides 2.88 kWh of electricity and 31 MJ of wood energies required in chitosan conversion. These figures only indicate the inputs of the processes and not the characterized impact on the environment specific to the defined industrial route. Direct kg-reagent-per-kg-chitosan comparison with enzymatic or fermentation routes remains difficult because published biological processes differ in enzyme or microbial inputs, carbon-source requirements, residual-protein removal, product yield, downstream purification, and targeted final purity. Some biological routes also retain a subsequent chemical-polishing step. Accordingly, the industrial HCl and NaOH inventory reported here should not be directly compared with non-harmonized laboratory-scale biological routes unless equivalent system boundaries, product specifications, and functional outputs are available.

This is applicable to all product sustainability claims, as any chitosan that is received by converters or formulators could contain significant and, therefore, mostly unobservable upstream burdens [122]. Leceta et al. [122] indicated the investigated chitosan-based film had lower effects in terms of carcinogens and fossil fuels, whereas it had higher effects in respiratory inorganics, land use, and minerals compared to the polypropylene reference film. Their assessment was based on the functional unit of 1 $m^2$ of packaging film and included material extraction, film production, and the end-of-life process.

The importance of chemical and utility requirements has been highlighted in a study conducted on industrial process modelling. Riofrio et al. [121] conducted a cradle to gate LCA of chitosan produced from Ecuadorian shrimp shells. They found that the impacts of global warming were 59.22 kg of $CO_2$-eq (per chitosan kg), the amount of freshwater consumed was 34.45 $m^3$ (per chitosan kg), and ecotoxicity and toxicity contributed significantly to all categories of impact [121]. Life cycle assessment studies have consistently identified the deacetylation stage as one of the most environmentally intensive steps in chitosan production because of the large quantities of sodium hydroxide required and the subsequent treatment of alkaline waste streams. While the use of crustacean shell waste provides a potential feedstock advantage, the environmental performance of chitosan remains highly sensitive to processing conditions. Consequently, sustainability outcomes are determined not only by the origin of the raw material but also by the intensity of chemical treatment, energy consumption, and waste management practices used during manufacturing [123]. The same cradle-to-gate assessment also demonstrates the importance of environmental indicators beyond climate change. Per kilogram of chitosan produced, terrestrial acidification was reported as 0.197 kg $SO_2$-eq, freshwater eutrophication as 0.0139 kg P-eq, and marine eutrophication as 0.00314 kg N-eq. Reported ecotoxicity impacts were 105.87 kg 1,4-DCB-eq for terrestrial ecotoxicity, 1.12 kg 1,4-DCB-eq for freshwater ecotoxicity, and 1.554 kg 1,4-DCB-eq for marine ecotoxicity [121]. These indicators demonstrate that environmental interpretation of chitosan production should not be restricted to GWP alone. However, they represent whole-system LCA results and should not automatically be attributed exclusively to acidic, alkaline, or saline effluent without supporting contribution analysis.

The findings from packaging-film life cycle assessments further reinforce the broader trend observed throughout chitosan production systems. Compared with conventional polypropylene-based packaging, chitosan films may provide environmental advantages in selected impact categories, particularly those associated with land use and certain emission-related indicators. However, these benefits are often accompanied by increased burdens in other categories because of the chemicals and processing requirements needed to transform chitosan into a functional packaging material. The production and use of reagents such as hydrochloric acid, acetic acid, and other processing chemicals contribute additional environmental impacts that are not immediately apparent when evaluating the material solely on the basis of its biological origin. As a result, the environmental profile of chitosan-based packaging is determined not only by the renewable nature of the feedstock but also by the intensity of extraction, purification, formulation, and film-manufacturing processes required to achieve the desired packaging performance [124].

The studies shown in Table 4 discover actual hotspots in the environment but cannot be directly compared. Muñoz et al. [18] used a consequential cradle-to-gate approach analyzing two industrial chitosan supply chains while packaging films involve all processes including production and end-of-life in case of necessity. The Indian supply chain analyzed in Muñoz et al. [18] report revealed that acid and alkali processes contribute heavily to environmental impact while thermal energy from coal and electricity play an important part in the European-related supply chain. The differences in results reflect inconsistency in production, energy supply, feedstock management and product quality instead of the environmental ranking of chitosan sourced from different locations. Therefore, the comparison will segregate inventory measurements from modelling results and reveal where the differences in product function or system boundaries prevent the numerical comparison.

**Table 4.** Quantitative evidence of hidden upstream burdens in chitosan-based packaging.

| **Material or product assessed** | **Region or supply chain** | **System boundary** | **Impact category (metric)** | **Quantitative result reported*** | **Dominant hotspot stage** | **Main hotspot input(s)** | **Key Takeaways** | **Ref** |
|---|---|---|---|---|---|---|---|---|
| Chitosan from crustacean shell waste | India | Cradle-to-gate | Climate change, kg $CO_2$-eq/kg | About 12 kg $CO_2$-eq per kg chitosan | Upstream chitin obtention and conversion | HCl production; chemical-intensive pretreatment | When production of acids and interim purification occurs, carbon multipliers do apply regardless of their waste-derived origin. | [18, 122] |
| Chitosan from crustacean shell waste | Europe-linked chain | Cradle-to-gate | Climate change, kg $CO_2$-eq/kg | About 77 kg $CO_2$-eq per kg chitosan | Energy-intensive upstream and distributed supply chain | Coal-based heat in China; electricity use across China and Europe | The fact that a material can carry the same label with regard to embodied carbon does not imply that they will all have the equivalent carbon/embodied energy values in different geographic regions or energy sources. | [18,125] |
| Chitin versus chitosan | India and Europe | Cradle-to-gate | Relative GHG burden | Chitosan shows about 1.6–1.7× higher GHG emissions than chitin | Deacetylation and downstream conversion | Additional alkali treatment and processing steps | The transformation of chitin to chitosan involves a lot of hidden costs that result from other than valorisation of waste. | [18, 94] |
| Industrial chitosan production from shrimp shell waste | Ecuador | Cradle-to-gate | Global warming potential | 59.22 kg $CO_2$-eq per kg chitosan | Raw material preparation and chemical processing | Shrimp-shell supply, NaOH, ethanol | Even before converting to packaging, an industrial facility typically has significant carbon impact potential if the raw material supply is conventional. | [121] |
| Industrial chitosan production from shrimp shell waste | Ecuador | Cradle-to-gate | Water consumption | 34.45 $m^3$ water per kg chitosan | Washing, purification, and process utilities | Process water demand and associated treatment | Prior to converting into final products, industrial plants may have used a large amount of water during interim processing, which adds further environmental pressures to potential carbon impacts. | [14, 121 122, 126] |
| Chitosan process with altered NaOH concentration | Ecuador, scenario comparison | Cradle-to-gate sensitivity analysis | Marine ecotoxicity change | About 17% reduction when NaOH concentration decreased from 50 wt% to 30 wt% | Deacetylation severity | NaOH concentration | The environmental performance of a biobased material may change once all of the emissions associated with the extraction chemicals used to produce the biobased product and the film-forming chemicals are included. | [121, 125, 127] |
| Chitosan-based film versus polypropylene film | Comparative packaging case | Comparative environmental assessment | Respiratory inorganics, land use, minerals | Chitosan-based films were higher than polypropylene in respiratory inorganics, land use, and minerals. | Raw-material extraction and film manufacture | HCl in extraction; acetic acid in film manufacture; glycerine for some impacts | A biobased film can lose its assumed environmental advantage once extraction chemicals and film-forming inputs are counted. | [94, 121] |
| Bio-based materials broadly | Cross-material review | Review of LCA literature | Cross-category environmental performance | No universal environmental superiority across all categories | Varies by product system | Processing energy, chemistry, transport, end-of-life assumptions | "Bio-based" should not be used as a proxy for low impact without life-cycle evidence. | [94, 122] |

***Note:** Outcomes correspond only to specific studies mentioned, hence they are not comparable in a direct way but rather as long as functional units, system limits, product features, feedstock processes, and life-cycle modeling assumptions are the same. Inputs for inventory assessment should be quite different from environmental risks. Thus, the results are mainly interpreted from the viewpoint of production hotspots and method variations, and not in terms of mere numerical values. For the Muñoz et al. [18] systems, the reported values were derived using consequential cradle-to-gate modelling with substitution to account for competing feedstock uses and by-product management rather than conventional mass- or economic-allocation rules. The India- and Europe-linked supply chains also

produced different chitosan grades for different applications and should therefore not be interpreted as a direct regional comparison of functionally equivalent products.

A detailed consequential LCA of two real industrial supply chains (Muñoz et al. [18]) illustrates this variability clearly. Chitosan produced in India from local shrimp shell waste showed approximately 12 kg $CO_2$-eq per kg, with chemical production (HCl and NaOH) and fertiliser use of protein sludge as key hotspots. In contrast, medical-grade chitosan produced via a global supply chain (crab shells from Canada, chitin production in China, deacetylation in Europe) reached approximately 77 kg $CO_2$-eq per kg which is driven primarily by coal-based heat in China and electricity consumption. The Indian chain benefited from substantial credits due to diversion of shells from animal feed (especially large water-use savings) and use of protein sludge as fertilizer, while the European chain gained credits from avoiding composting emissions (particularly ammonia). Indirect land-use change was relevant only for the Indian supply chain. These results demonstrate that even within the same material family, environmental performance can differ by nearly an order of magnitude depending on raw material source, processing location, energy mix, and by-product management.

### *3.3. Additive selection: performance enhancement* vs *end-of-life uncertainty*

For chitosan packaging, the use of additives is not intended as a replacement for plastic but rather to compensate for the natural limitations associated with chitosan, such as its sensitivity to moisture, limited strength, and poor barrier properties with respect to gases. The critical window for selecting appropriate additives occurs at this point within the life cycle of the material, as the short-term benefits associated with the performance of a package in the present may generate significant uncertainty with regard to the environmental consequences of its use in the future. While additives may constitute only a minor component (a few percent) of the total mass of a material, the characteristics of those additives significantly affect its degradation, persistence, and fate in the environment following disposal. Therefore, in this section, three common formulation approaches (chitosan and a biodegradable additive, chitosan with nanoparticles, and functional additives with conventional petroleum-based plastics) are compared to demonstrate how the selection of an additive affects the properties of the final product for comparative use and disposal rather than simply categorizing the materials as being better or worse by nature.

#### 3.3.1. Chitosan with biodegradable additives

Chitosan films, which possess unique properties, can be enhanced with either fully biodegradable or conditionally biodegradable materials. The use of fully biodegradable or conditionally biodegradable additives such as polyvinyl alcohol (PVA), starch derivatives, cellulose nanofibers, or natural plasticizers can improve certain functional properties of the chitosan film. More specifically, these additives help mitigate some of the limitations associated with chitosan materials, including brittleness, sensitivity to moisture, and lack of mechanical strength, by modifying the polymer chain mobility and intermolecular interactions within the film matrix [128].

These additives may improve film performance and contribute to food preservation; however, reductions in spoilage and food waste must be demonstrated under relevant storage conditions. Moreover, the improvements in oxygen barrier properties associated with biodegradable polymer blends slow the oxidative degradation of food products, while the improvements in flexibility and structural integrity reduce the likelihood of cracking, tearing, and package failure during handling and storage. The literature has demonstrated that the blending of chitosan films with biodegradable polymers results in reduced oxidation rates and reduced bacterial growth, particularly with respect to the packaging of fresh produce and meat [73, 129].

End-of-life sustainability behavior for these blends of natural polymers is not based purely on chitosan as a single polymer component but rather on a combination of multiple polymers used in the formulation of composite blends. Each component has its own distinct degradation pathway, which is not typically synchronized under actual disposal conditions, leading to significantly different rates of degradation for different components of the composite blend in compost or soil. For example, although chitosan typically breaks down quickly, other additives (such as polyvinyl alcohol or chemically modified starch) may be able to decompose completely only under specific industrial composting conditions [130-132]. As a result, when degradation occurs incompletely (particularly for highly crosslinked polymer networks or for polymeric materials formulated with excessive amounts of additives), persistent polymer fragments can remain.

If additives (such as chemically crosslinked polymers or excessive loads of polymers) are present in composites that make them structurally durable long term during their use phase, then their crosslinks may inhibit microbial access to those additives, thereby reducing the rate of their decomposition after disposal. Therefore, while some of the materials may be functionally biodegradable when considered in theoretical terms, they cannot meet conservative definitions of biodegradability under actual conditions during composting or where organic debris is decomposing. An example of such a loss of composting efficiency could occur after partial degradation has occurred. Partial degradation typically affects the total water-holding capacity of the compost and the total microbial community activity present in the compost and can inhibit the quality of the compost when undecomposed polymers remain in the finished product [133, 134].

The compatibility of biodegradable additives as they degrade is part of the environmental performance of a chitosan packaging system. Ultimately, whether each of the materials is referenced individually as a biodegradability additive does not matter; rather, it is the ability to degrade as a total package in a coherent manner when disposed in a manner similar to that of the average consumer that will determine the sustainability performance of chitosan packaging. Because of this, choosing additives is an important life cycle decision rather than simply a minor formulation consideration [134].

### 3.3.2. Chitosan with inorganic nanoparticles

Chitosan films typically incorporate inorganic nanoparticles such as silica, zinc oxide, titanium dioxide, and silver to provide chitosan films with mechanical reinforcement, UV shielding, improved barrier performance and antimicrobial characteristics. Inorganic nanoparticles can complement or overcome the limitations of chitosan when blended with polymers alone, especially in applications that require active antimicrobial or optical properties. In addition, the use of these additives at low concentrations (often < 5 wt%) can significantly affect both film performance and the presence of preservatives [134]. Nanocomposite chitosan films can increase the shelf life while reducing waste by limiting spoilage caused by microorganisms or oxidative deterioration. The enhanced antibacterial activity of the nanocomposite chitosan films can also reduce surface contamination, and improved ultraviolet and oxygen barrier properties can inhibit the photodegradation and oxidative spoilage of packaged foods. Understandably, nanocomposite chitosan films will appear to be valuable and likely sustainable packaging alternatives when considered in terms of use-phase performance.

As is the case with biodegradable additives, the key trade-offs are most apparent at the end of life. However, the nature of the tradeoff differs. Upon composting or leaving the chitosan matrix to degrade under soil conditions, the biopolymer matrix should undergo relatively rapid degradation; however, inorganic nanoparticles do not follow the same degradation pathway. Therefore, a decoupling between matrix degradation and the persistence of the added inorganic nanoparticles occurs. Because of this decoupling, as the biopolymer matrix degrades, nanoparticles can be released into soil or aquatic environments; therefore, their fate, movement, and bioavailability in the environment are unpredictable. When nanoparticles are released into an environmental matrix, they may aggregate, undergo surface transformations, and/or interact with organic matter, complicating the evaluation of their long-term environmental behavior. For example, it has been well documented that certain metal oxide nanoparticles accumulate in soil environments and may affect the structure of microbial communities or nutrient cycling processes [135, 136]

In contrast to their relatively small weight, inorganic nanoparticles are likely to play an important role in shaping the long-term environmental fate of a composite material. While the bulk chitosan matrix may be involved in its natural process of degradation, distinct environmental stressors related to the presence of nanoparticles still exist. In this regard, the nature of a composite material evolves from being biodegradable to being an extended-release pathway for the diffuse release of nanoparticles. Once released, nanoparticles may be transported beyond the original disposal location through soil, water, or sediment pathways, complicating efforts to identify and attribute long-term environmental impacts to a specific product source. The biodegradability of the matrix alone does not reduce the risk associated with the presence of nanoparticles; in fact, the biodegradability of the matrix may very well lead to enhanced release of nanoparticles because of the breakdown of the bulk host material in which the nanoparticles were originally contained [137-140].

### 3.3.3. Conventional polymers with additives

Petroleum-based conventional packaging often has been modified to improve food shelf life and durability by incorporating additives such as antioxidants, plasticizers, UV stabilizers, and antimicrobial agents. These additives increase the ability to resist mechanical fatigue, ultraviolet damage, and oxidation through the packaging supply chain (transportation and storage). Additionally, these materials typically exhibit high barrier properties and long shelf life, which can help reduce food waste caused by wasted products under challenging application conditions [141-143].

Unfortunately, these polymers are nonbiodegradable, meaning that they typically just accumulate in landfills, persist in the environment, and remain in nature for long periods of time after their useful life. Additionally, because the additives are encapsulated within a relatively stable polymeric structure, the behavior of the additives in such applications is usually more predictable than that in biodegradable matrices. In contrast, the additives may be released quickly because of the decomposition of the biodegradable matrix; they typically leach out of the intact host material more slowly [8, 144].

The analysis indicates that there is a significant tradeoff between biodegradable matrices containing long-lived additives, which produce substantial uncertainty with respect to the future environmental fate of the material, and fully long-lived, stable formulations (continuously strong additive migration potential). Biodegradable materials do not eliminate the potential for adverse environmental effects; in some instances, these materials accelerate the movement of the additive from the host matrix into the environment [141, 145, 146].

### 3.3.4. Cross-stage interpretation

In each of the three systems, the additive-selection process is an excellent example of how the choices made in performance-based materials influence sustainability throughout their entire life cycle. Additives are generally selected to address functional limitations of materials during use, and the selection of additives can also significantly impact how materials behave environmentally after the end of the use period. For instance, materials that provide enhanced barrier properties and extended shelf life will benefit from use, but their degradation will also result in various environmental impacts much later. In the absence of an assessment that considers both the performance improvements realized through the use of the additive and the long-term potential impact of those additives on the environment, there is a good chance that no consideration was given to the trade-off between short-term performance gains and long-term impacts on the environment [73, 147, 148].

For example, while it may be true that chitosan is biodegradable, it may not be safe for the environment unless the chemical structure or formulation of the additive used to enhance performance is compatible with the expected degradation of the resin. Notably, even though the total volume of additives used in a material could represent a small percentage of the total weight, their chemical composition and subsequent degradation will likely govern how that material interacts with soil, water and waste management systems after it is disposed of. Therefore, an assessment focused only on the use of renewable feedstocks or nominal biodegradability will not capture the factors that ultimately govern the environmental results of that material. In the absence of an assessment that considers both additive residue and the expected degradation path, the potential for overstating the environmental advantages of the material or obscuring other important end-of-life trade-offs exists [148]. The environmental significance of these additives extends beyond improvements in mechanical or barrier properties. By reducing microbial growth, oxygen permeability, moisture transfer, and oxidation reactions, additive-enhanced chitosan films can extend the shelf life of food products and reduce food losses during storage and distribution. Consequently, the sustainability contribution of additives must be evaluated not only through their production and disposal burdens but also through their capacity to reduce food waste throughout the use phase of the packaging system.

### *3.4. Use-phase trade-offs: food waste reduction vs rebound consumption effects*

Although the use phase advantages of high-performance packaging solutions are commonly mentioned, they tend to be overlooked and underexamined stages of a food package's life cycle. Shelf life and food preservation have been cited as environmental advantages of these packages; however, these advantages may be countered by consumer behavior and system responses that can result in diminished or offsetting sustainability results. Chitosan-based films that contain antibacterial and antioxidant additives can extend the shelf life of perishable goods by inhibiting microbial growth and slowing oxidation reactions. Consequently, packages with extended shelf life will experience less deterioration during storage, transportation, and retail display, which will lower the probability of wasting packaged food before it is consumed [148, 149].

The reduction in food waste due to packaging with extended shelf life can also reduce upstream emissions from agricultural and supply chain processes associated with the production of food lost to spoilage. A reduction in food waste can reduce the land use burden from livestock production/feeding, fertilizer application, irrigation, processing and cold chain logistics. Many proponents of active and nanocomposite packaging technologies cite the avoided impacts of food being wasted as a significant argument in favor of these technologies. In addition, it is important to remember that these avoided impacts are indirect versus direct and occur if the longer shelf life of food equals a reduction in food waste rather than simply changes to the purchasing behavior or storage and eating behavior of consumers [148].

Enhanced packaging performance might sometimes cause rebound effects that negate use-phase gains. Improved durability, hygiene, and shelf life frequently reinforce single-use consumption habits by promoting disposability and convenience-oriented behaviors. When packaging is regarded as "safer" or "cleaner," consumers may be less likely to reuse containers or reduce packaging demand, even if the material is biodegradable. Improved performance in chitosan-based packaging may justify its expansion into previously unpackaged or minimally packaged applications, increasing overall packing throughput. As a result, total material flows might increase while the per-unit environmental effect decreases [150-152].

Shelf life extension can also lead to higher stocks, longer distribution chains, and a greater reliance on individually packaged portions. While each unit of food may generate less trash, the total amount of packaging entering the system may increase. This effect is especially important for biodegradable materials, as lower perceived environmental danger might reduce incentives for material reduction. Life-cycle studies are increasingly demonstrating that material substitution alone does not ensure that the level of environmental benefit is overestimated when total consumption increases. In such circumstances, savings from reduced food waste may be countered by higher packaging production, processing energy, and waste management demand [149, 153].

### *3.5. End-of-life pathways: biodegradability vs environmental realism*

Chitosan-based polymers used as an alternative form of food packaging may satisfy regulatory requirements for biodegradability because of their inherent chemical composition (biodegradable and antimicrobial). However, how a chitosan-based polymer will biodegrade at its end of life is dependent upon several additional factors other than just the presence of a biodegradable backbone. In addition to the presence of a biodegradable polymeric backbone, the rate and extent to which chitosan biodegrades are based upon intrinsic properties such as molecular weight and degree of acetylation as well as extrinsic environmental conditions. As such, the results from biodegradability tests conducted under controlled conditions do not always equate to the results from testing conducted at real-world disposal sites [134, 153].

Whenever the biodegradability of chitosan is evaluated at the end of its useful life, all the conditions related to the ability of the polymer to biodegrade, the conditions under which the polymer can biodegrade, and any residual materials left from the polymer must be considered when assessing the potential risk to the environment. Additionally, environmental fate considerations for any chitosan-based film include not only the chitosan polymer but also other components added during manufacturing, such as additives, residual solvents, nanoparticles, and other ancillary materials, that may behave differently after disposal. Thus, the primary concerns when evaluating the biodegradability of chitosan can be defined as the relationship between the conditions needed to achieve biodegradation, the byproducts formed as a result, and how those byproducts impact ecological systems upon exposure to the byproducts. Therefore, the environment may be subjected to risk through the introduction of chemical substances prior to the ultimate breakdown of the chemical substance unless appropriate risk assessments are completed prior to material introduction [119]. **Figure 5** summarizes the major end-of-life pathways for chitosan-based food packaging and the uncertainties associated with each route. While industrial composting can facilitate rapid biodegradation under controlled conditions, landfill disposal and environmental leakage often result in slower degradation, greater persistence, and increased uncertainty regarding the fate of additives and nanoparticles. The framework highlights that environmental outcomes are determined not only by material biodegradability but also by disposal pathway, formulation design, and local environmental conditions.

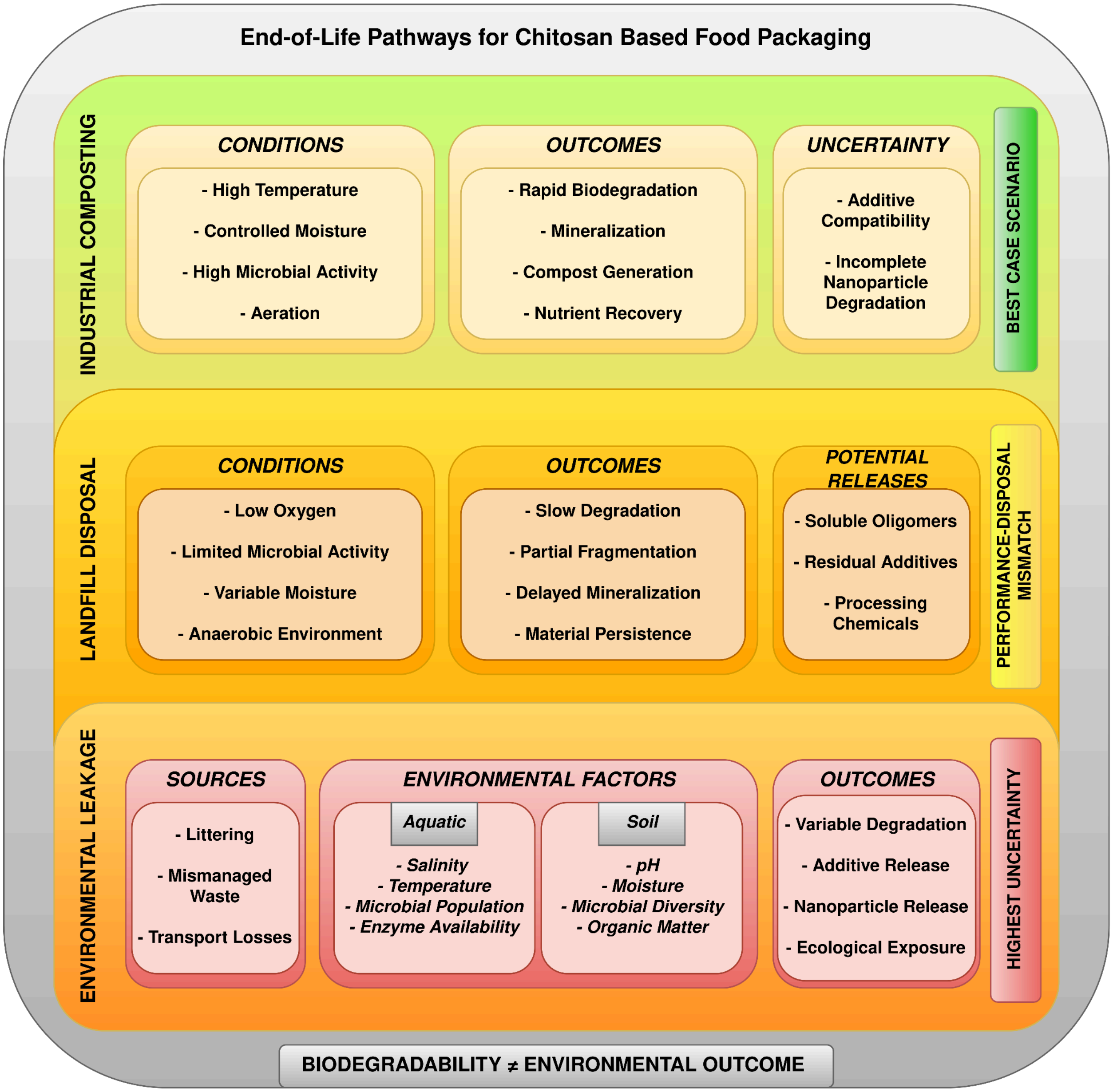

**Figure 5. End-of-life pathways and environmental uncertainties associated with chitosan-based food packaging.** *Industrial composting represents the most favorable disposal scenario because controlled temperature, moisture, aeration, and microbial activity facilitate biodegradation and nutrient recovery. In contrast, landfill disposal and environmental leakage introduce conditions that may slow degradation, promote persistence of additives or nanoparticles, and increase uncertainty regarding long-term environmental fate. The figure highlights that biodegradability alone does not determine sustainability outcomes, as environmental performance depends on disposal pathway, formulation composition, and local environmental conditions. The degradation and residue pathways shown are*

*formulation-dependent and should not be interpreted as equivalent for neat chitosan, polymer blends, crosslinked films, and inorganic nanocomposites.*

Composting on an industrial scale is considered one of the best methods for the disposal of chitosan-based film materials and accelerates the hydrolysis and breakdown processes because of the high temperatures, controlled moisture levels and high number and activity of microorganisms. Because of the potential benefits from the composting of these materials, composting can be viewed as the optimum or best-case scenario. Unfortunately, the availability of composting capacity is not necessarily universal, and several types of materials are poorly suited for or do not complete their composting journey. Chitosan film materials often have additive materials that change the rate of degradation of the film versus the matrix. Even though the addition of an additive will increase the performance during use, those same additives may not degrade as quickly as the chitosan matrix does. Therefore, it cannot be assumed that if the full material system is exposed to the composting environment, it will fully mineralize [16].

On the other hand, the often low temperatures, limited microbial activity, and reduced oxygen availability found in landfills often characterize the conditions of landfill disposal. Under these conditions, chitosan biodegradation is significantly slowed. Anaerobic landfill environments slow the enzyme pathways responsible for the biodegradation of chitosan. Partial degradation may occur; moreover, owing to the restriction of microbial access to the polymer surface, degradation may occur over extended timescales, and the complete mineralization of chitosan into nontoxic mineral compounds is unlikely. Instead, chitosan may persist within the environment as slowly fragmenting and degrading solids. This challenges the idea that the use of the biopolymer chitosan within food packaging inherently helps avoid long-term environmental accumulation and reduces the environmental impact [154].

Additionally, as the material ages and biodegrades, the residual processing chemicals and particles, additives, and soluble oligomers may slowly leach into surrounding areas and pollute the soil and groundwater. Chitosan-based food packaging cannot be individually assessed without considering the effects of the additives on the environment. Landfill environments are mostly anaerobic, especially after the initial aerobic phase; therefore, as a result, additives (synthetic or natural) accumulate rather than being fully mineralized. The presence of additives complicates the biocompatibility of the chitosan film. Although some additives, such as blending chitosan with synthetic polymers or nanoparticles, slow the biodegradation rate of the chitosan film, some additives can help facilitate biodegradation. Furthermore, some additives, including plasticizers such as glycerol, provide a carbon source for microorganisms, which may increase biodegradation rates because the additive enhances certain properties of the chitosan film; these properties are dependent on the type of additive utilized [155]. These additives may help facilitate the decomposition of chitosan-based films in more anaerobic environments, such as landfills and aquatic environments subjected to improperly treated chitosan film materials.

One route by which food packaging materials may enter the environment is by improperly disposing of them or through a lack of sufficient waste management strategies, which represents a significant nonideal end-of-life option for food-packaging materials in aquatic environments. Biodegradable biopolymers (i.e., chitosan) are often assumed to be environmentally safe if they are present in water; however, this assumption is not always correct. While chitosan may degrade or disintegrate under some aquatic test conditions, the extent of degradation is highly dependent on temperature, the number of microorganisms present, salinity, and whether enzymes are available to assist in degradation, but none of these conditions will likely match those under controlled laboratory conditions [156].

Under these conditions, chitosan films can either fragment or fully mineralize, and any additives or residual processing chemicals incorporated into the film can also be released into the aquatic system. Concerns about ecological exposure exist because of the presence of nanoparticles, insoluble additives/materials, and soluble residues. Chitosan has antimicrobial properties that may help it to inhibit microbial colonization when used, but in some cases, it may further inhibit microbial colonization after disposal and slow the degradation rate of chitosan in the environment. Careful formulation of chitosan films is also critical because the use of additives (such as plasticizers and barrier modifiers) can substantially change the hydrophilicity, structure, and microbial access to chitosan films, thereby affecting their degradation in the aquatic environment [96].

## 4. Understanding Trade-offs: Where Sustainability Narratives Break

The sustainability profile of chitosan-based packaging is determined by the way it has been sourced, processed, formulated, utilized, and disposed of. Specifically, variations in the extraction, synthesis, and end-of-life methods used can considerably influence not only the functional properties of chitosan as a biodegradable material but also the total environmental impact associated with chitosan as a biodegradable material. **Figure 6** synthesizes the major sustainability trade-offs that emerge across the life cycle of chitosan-based food packaging systems. While chitosan is frequently promoted as a sustainable material because of its waste-derived origin, biodegradability, and food-preservation capabilities, each stage of its production and use introduces environmental burdens that may offset these benefits. The framework illustrates how increasing process intensity often improves material purity, functionality, and packaging performance, but simultaneously increases chemical consumption, energy demand, wastewater generation, and end-of-life complexity. Consequently, the sustainability of chitosan-based packaging cannot be assessed through biodegradability alone and must instead be evaluated through the cumulative interactions between feedstock sourcing, extraction methods, manufacturing practices, use-phase benefits, and disposal pathways.

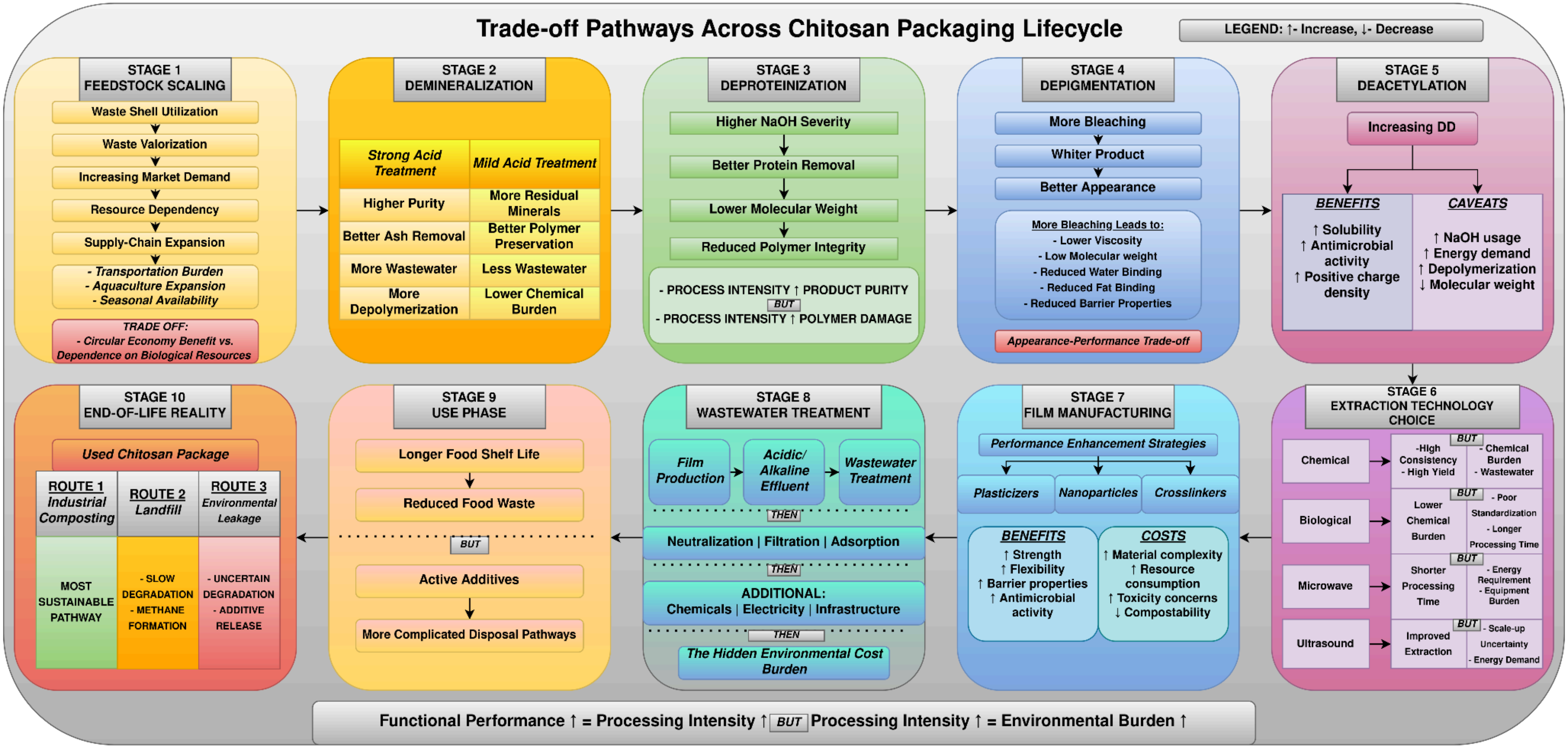

**Figure 6. Trade-off pathways across the chitosan packaging lifecycle.** *This conceptual framework synthesizes data available in existing literature on the inputs used for chitosan production, composition- and production-dependent performance characteristics of packing, and disposal uncertainties. The industrial life cycle analysis highlights the role of the manufacturing stage in India as a key contributor to the chitosan supply chain, and the role of thermal energy and power consumption in the European system [18]. This evidence illustrates different product systems and is not meant for direct regional comparison. The figure shows relationships between improved packaging performance characteristics and processing requirements, efficiency of food preservation, and disposal methods.*

### *4.1 Scale-up and feedstock dependency*

At the stage of scaling up production, one of the most significant trade-offs occurs. Chitosan is often referred to as a waste-derived product; however, the sustainability of chitosan is dependent on whether shell waste remains solely a secondary byproduct or if it is viewed as an economically viable feedstock with greater demand. Currently, chitosan is used in the biomedical and pharmaceutical fields and several other sectors; therefore, the extension of chitosan to food packaging will place additional demand on the resources used to produce chitin. Owing to the magnitude of the global packaging market versus the amount of chitin that is produced today, even a small percentage of conventional plastic materials replaced with chitosan will require a significant increase in chitin output [157].

This creates an overarching systems-level issue. If chitin is derived from shells and has economic value, then chitin derived from crustacean shell waste will likely no longer be considered a low-burden residual stream from an economic perspective. The supply chain for chitin depends on the availability of shells (seasonal availability), logistical supply chain issues, aquaculture expansion, and/or the expansion of marine resources harvested. Furthermore, the waste-valorization narrative of using chitin as a raw material will be undermined, and there is potential for broad-scale negative ecological consequences, particularly with respect to the importance of crustaceans in aquatic nutrient cycling or sediment turnover, and issues related to aquaculture expansion, such as habitat loss, greenhouse gas emissions, water pollution, and increased disease risk [157, 158].

Increased dependence on chitin-based packaging can add to transportation-related burdens if the handling of shell waste, extracted chitin, processed chitosan and completed film products occurs in separate locations. Additional logistical costs would likely increase the total carbon footprint of the system, potentially becoming even more significant as supply chains become larger in terms of geography or time. Transport must be evaluated alongside extraction and production when determining the total environmental burden of chitosan packaging [125, 159].

### *4.2 Extraction conditions and quality–sustainability trade-offs*

Chitin extraction involves a quality-sustainability trade-off: stronger acid/base treatments can reduce ash and protein residues, but they may increase chemical burden, wastewater load, and polymer depolymerization.  Most commercial producers use strong acids to extract solvents from calcium carbonate and similar mineral components from chitin shells. Demineralization of chitin can be achieved using chemical demineralization methods, where the process is highly dependent on the type of chitin shell, the strength of the acid, and the temperature and length of exposure to the acid. Depending on the conditions used in the extraction process, the quantity of residual ash remaining after the process is affected by the degree to which severe extraction conditions are used. In turn, increased levels of calcium carbonate and other minerals increase the level of depolymerization and structural and functional quality of chitin. In comparison, milder acids will allow for the preservation of the chitin through lower amounts of polymer degradation and an increase in residual minerals. In any case, even though most allomorphs exhibit identical levels of crystallinity, the manner in which amorphous regions are preferentially removed during chemical demineralization will cause differences in polymer integrity. Thus, increased levels of purification of demineralized chitin are not necessarily related to greater sustainability or functional superiority; they result from more severe extraction conditions and provide a greater likelihood of quality loss [160].

There will always be a compromise on the effectiveness of deproteinization by chemical means. The dependent factors affecting deproteinization are the type of base being used; temperature; time; and the raw material that is being used; however, the use of NaOH is preferred in industry but has a limited ability because of partial depolymerization and hydrolytic degradation of the chitin structure. The increased protein removal under process conditions may, in some cases, lead to a reduction in the molecular weight of the downstream chitosan products and thereby affect the performance of the resulting chitosan. Once again, process intensity versus product performance cannot be optimized independently [121].

Depigmentation creates another compromise that exists in the performance versus visual characteristics of a product. Since many variables are related to the type of raw material that is being used in an extraction and the conditions used for extraction, a number of different residual colors can be produced from shell fishery byproducts; thus, the conditions being used to bleach (depigment) these feedstocks vary from feedstock to feedstock and from batch to batch. However, excessive aggressive depigmentation may damage the chitin matrix, thereby lowering its viscosity, decreasing its molecular weight and reducing its water and fat binding capabilities [161]. If any of these effects occur, this will have an adverse effect on the barrier properties of the resultant products made from food packaging. Thus, depigmentation should not be viewed as being beneficial universally; rather, its value or economic use is dependent on whether the improvement in appearance outweighs the added chemical burden as well as the possible loss of functional properties [162].

Deacetylation is another point of intersection between processing intensity and production goals. In general, increasing the DD produces materials that are more soluble in dilute acids and possess a greater degree of cationic behavior associated with antimicrobial properties, making these materials ideal candidates for use in food packaging applications. Conversely, obtaining very high DD levels often requires more vigorous alkaline processing, which can lead to increased levels of depolymerization, resulting in lower molecular weights. If the alkaline treatment is too mild, the resulting material may still possess a significant amount of chitin-type characteristics and therefore will not have the desired combination of solubility and activity. Therefore, chitosans obtained via commercial means have a wide variety of DDs because the severity of the processing method needs to be balanced with the application of the material [100, 163].

Biological processing options (such as lactic acid fermentation) provide a potentially less chemically intensive method; however, they also present their own unique challenges for optimization. In particular, the characteristics of the resulting product depend on many factors, including the pH, temperature, substrate composition, salt concentration, length of incubation, and microorganism strain used to produce. These types of interactions affect the extent of demineralization and deproteinization (i.e., achieving an equivalent level of product consistency) and help to illustrate why using biological means for extraction continues to be advantageous from an environmental perspective yet difficult to standardize on a larger scale [15].

Physically assisted extraction techniques have frequently been cited as viable alternatives to reduce extraction times and increase yield or product quality. Microwave-assisted, ultrasonic-assisted, and use of supercritical carbon dioxide and hot water extractions may accelerate demineralization or deproteinization rates and, in some cases, be combined with biological methods to increase effectiveness. Nevertheless, the assumption of an inherent sustainability advantage should not be made. Any reduction in either chemical or processing time still needs to be considered in consideration of the equipment requirements, amount of energy utilized, and scalability of the processes. Therefore, these extraction techniques should more appropriately be viewed as promising process intensification approaches rather than as recognized methods that address the larger sustainability trade-off [130].

### *4.3 Manufacturing burdens and waste treatment realities*

There are a significant number of trade-offs associated with how films manufactured with chitosan can be differentiated as a result of the production process and the additives used to produce the film. Some methods used to enhance film characteristics (such as flexibility, strength, barrier properties and antimicrobial properties) may do so at the expense of materials, energy and the environment. Ultimately, the characteristics of the film produced will be determined by the decisions made in the formulation of the film during production, including how the film will be used after production (i.e., its entire life cycle, not just the time spent at the factory) [73, 132].

Production methods such as solvent casting, solution blending, and electrospinning typically require many chemicals to process and generate large quantities of wastewater, as well as a large quantity of solvents that need to be handled as a result of using those production techniques. Both the purification process and film-making processes of chitosan create large amounts of acidic and alkaline wastewater. If not treated properly, these waste streams can lead to soil and water contamination and create potential risks for workers or the public because of their corrosive nature. Because of these characteristics, solvent selection and waste management are sustainability issues rather than simple operational issues [164-167].

The amount of waste produced from chitosan is directly related to the chemistry used to manufacture the films. Chitosan cannot be dissolved in neutral or basic media; therefore, chitosan is usually dissolved in acidic solutions that protonate amino groups or disrupt intermolecular bonds that hold the crystalline structure of chitosan together. Through this process, a film-forming solution can be created; however, after this step, the film is dependent upon several steps to be properly manufactured, including washing, neutralization, and drying. Residual acid from the washing process can alter some properties of the film, such as the network structure, plasticity and manner of crosslinking under certain conditions [102, 130, 166].

Manufacturing processes are supported by various types of activities. A hidden but large part of the total manufacturing footprint is wastewater treatment [167]. Neutralizing, filtering, adsorbing, and related treatment activities require additional chemicals, electrical power, and infrastructure, which increase both the economic and environmental costs of the manufacturing activity [18]. The requirements to treat effluents that have high levels of dissolved salts, suspended solids, organic materials, or trace contaminants are even more intensive; therefore, any potential sustainability benefit from biobased films may be significantly diluted because of the additional chemical and energy requirements needed for solvent management and effluent treatment.

## 5. Comparative Analysis: Chitosan vs. Other Packaging Materials

### *5.1. Comparative framing across the life cycle*

Comparing the environmental performance of chitosan as a food packaging material with that of other food packaging materials is insufficient for considering isolated materials and the physical characteristics of the materials. Many people refer to packaging materials as “biobased,” “biodegradable,” or “recyclable” in terms of their environmental impact. However, these classifications do not establish whether a particular packaging system is more environmentally superior to other packaging systems from a life-cycle perspective.

When considering the total environmental impact of a packaging system from its own production to the point in time when the package is discarded, this provides the best measure to determine how three different packaging systems compare with each other from an environmentally friendly perspective. During the process of conducting a life cycle assessment (LCA) of a packaging system, it has been continuously demonstrated that there is no packaging material that has the lowest environmental burden (such as climate change emissions, energy consumption, water consumption, toxicity, land use, and waste disposal) across every environmental category. The evaluation of the total environmental burden associated with a packaging system changes depending on the functional unit selected when the LCA is being conducted, the system boundary used when the assessment is being conducted and the assumptions made regarding the final disposal of the product [168].

Chitosan should not be understood as a staple substitution for traditional types of packaging by some consumers; rather, its attraction lies in certain characteristics, including possible valorization of shell waste feedstock, the natural antimicrobial properties of chitosan, its ability to create films, and its potential for active packaging/coating applications. However, these characteristics also have disadvantages, such as moisture sensitivity and low mechanical strength; the need for chemical-intensive methods of purifying chitosan; and questions regarding the end-of-life disposition of chitosan if it has been blended with or contains additives. Thus, a fair comparison of chitosan to PLA, PET and glass should consider not only where chitosan appears to have advantages but also where its sustainability profile declines when it is considered as part of an overall life cycle approach [169].

In order to achieve a functional comparison, we need to compare packaging systems based on the ability to protect the quantity and type of food during a particular storage and distribution period. Important parameters in this respect include packaging format, wall thickness, mechanical stability, the quality of oxygen and water vapor barrier, and practical preservation effectiveness. In cases when protection involving chitosan should utilize blends, plasticizers, crosslinkers, and nanofillers, the relative contributions of these additives should also be considered in making the comparison.

### *5.2. Feedstock origin and upstream resource dependency*

From a feedstock perspective, chitosan and polylactic acid (PLA) offer opportunities for renewable-resource utilization, whereas polyethylene terephthalate (PET) and glass are generally produced from fossil-derived and mineral feedstocks, respectively. However, feedstock renewability alone does not establish lower life-cycle environmental impacts. Compared with products derived from virgin fossil or mineral resources, chitosan, in particular, has the potential for valorizing a byproduct stream when produced from crustacean shell waste. In addition, PLA is also considered to be renewable because it is produced from agricultural feedstocks that are high in carbohydrates (generally from corn or sugar), which are converted to lactic acid prior to the production of PLA. On the other hand, PET is produced using fossil-derived petrochemical products, and glass is produced by extracting minerals and using a high amount of energy to process them [170].

Nevertheless, a critical review of the relevant literature is needed. The upstream advantage of chitosan is highly dependent on the fate of shell waste; whether or not it is a truly low-burden residual stream or becomes a strategically demanded feedstock as industrial-scale production expands. Substantial demand for chitosan may arise in food packaging, and if so, the current narrative regarding the waste valorization of chitosan could break down because more shell materials will need to be collected, transported, purified, and managed through more formalized supply chains. This does not inherently mean that shell waste will be exploited; however, the environmental credit associated with "waste" becomes increasingly prone to sensitivity on the basis of allocation assumptions, regional seafood-processing patterns, and transport intensity. As a result of these various factors, the feedstock advantage of chitosan may exist; however, it is conditional rather than unconditional in nature [168].

PLA faces a different set of upstream trade-offs than other plastics do. Although it has a renewable source of feedstock and thus does not rely on fossil feedstocks, it does incur agricultural burdens related to growing crops for raw materials, which include land use, fertilizers, irrigation, and the potential for eutrophication. Life cycle assessments (LCAs) conducted on the PLA suggest that these agricultural burdens can be substantial; therefore, the environmental benefits of the PLA may be less significant than assumed on the basis of its renewable nature. Consequently, the upstream environmental burden of the PLA does not eliminate fossil fuel extraction but rather shifts it from fossil material extraction to agriculture. Conversely, while PET and glass are both based upon the nonrenewable extraction of resources, they are part of a well-developed and mature industrial supply chain, thus making the environmental impacts of both types of plastics much easier to determine. Thus, chitosan would appear to provide an upstream advantage only when adequate shell-derived feedstocks exist with minimal increases in resource use impacts and/or logistical complexities [171].

### *5.3. Production-stage burdens and process intensity*

The sustainability narrative surrounding chitosan production is complicated by the fact that it is not harvested and then formed into packaging but is instead harvested using multiple physical and chemical methods to extract (demineralization, deproteinization, and deacetylation) and process (washing and drying) chitosan, all of which require the use of acid, base, water, and thermal energy and produce both saline and chemically loaded waste effluent. Therefore, the environmental attributes of chitosan are determined by both its renewable source and the intensity of production of the material (including the method of production, as it relates to the sustainability of chitosan). The label "biopolymer" does not inherently imply that the production of the material will be of lower impact than the production of nonorganic polymers [172].

PLA, like conventional plastics, also requires considerable amounts of resources to manufacture. The production process for PLA includes fermentation, purification and polymerization (i.e., conversion into a polymer form). Each step of the production process can require a significant amount of energy. The environmental impact of PLA and conventional plastics can vary widely depending on how each step of the production process is modeled and how waste from both types of materials is treated after disposal. For example, LCA studies that compared PLA to PET milk bottles revealed that PLA was not considered advantageous for using PET (under current disposal options) until future disposal options were made more favorable. Importantly, this finding demonstrates that the use of renewable sources for feedstocks does not necessarily indicate that the overall environmental impact of production and disposal will be less than the impact of the production and disposal of conventional plastics [173].

While PET production remains fundamentally linked to the use of fossil-based energy sources and petrochemicals as raw materials, it is highly efficient and has demonstrated a high degree of industrial maturity. This industrial maturity creates the potential for mature polymers produced from fossil sources to have efficient scale operating characteristics. For example, the 2024 LCA study conducted on PET, PLA, and chitin nanocrystal-reinforced PLA provides further insight into this issue. The production-stage climate impact associated with PET was determined to be 3.21 kg $CO_2$-eq, which is greater than the 3.04 kg $CO_2$eq of SSC (specific surface area) and its variants with a chitin nanocrystal/PLA composite reinforced with 5 wt% chitin (4.26 kg $CO_2$-eq). The finding that the use of chitin reinforcement did not mitigate the overall climate impact from the production of that material indicates that the additional burdens associated with the production of chitin represented an increased climate footprint. As a result, the authors of this report identified hydrochloric acid consumption, water demands during hydrolysis, and energy required for drying as hotspots associated with the chitin component and noted that reductions in those environmental indicators can significantly reduce the climate impact associated with the production of chitin nanocrystals. The information contained in these findings strengthens the position of a critical review as to whether or not the use of materials derived from chitin is an important issue; rather, it strengthens the position that the sustainability of processes for producing these materials must meet acceptable standards [174].

The economic value of glass differs from that of other materials because of the significant energy used to produce it through melting and shaping. Life cycle assessment (LCA) studies of hollow glass packaging regularly use three key variables to determine its environmental performance: furnace energy use, package mass, and percentage of recycled content. Glass has the potential to improve dramatically by reducing the material content and increasing the use of recycled glass, but there will always be a high energy demand associated with producing it within the production phase, unless these improvements occur. In summary, each of the four materials shifts the burden of production to a different production domain: chitosan (chemical) to water or effluent treatment; PLA (from fermentation to polymer) to reverse osmosis (RO); PET (primary feedstock) to fossil fuels; and glass between thermal energy and mass. None of the four domains are entirely free of production burdens [174, 175].

### *5.4. Functional performance and packaging suitability*

The number of potential products for chitosan, compared to products made from glass, PLA, or PET, is limited largely because of the physical properties of chitosan itself. This generally means that pure chitosan cannot be used to completely replace rigid or liquid packaging materials that have been approved for use today because of its ability to absorb moisture [132]. Pure chitosan generally does not possess the dimensional stability, structural integrity, or moisture barrier properties needed to meet the performance requirements of many of today's more advanced packaging applications. Compared with traditional packaging materials, pure chitosan can offer many more types of functionality because it possesses functional properties that allow it to provide unique functions that traditional packaging materials do not possess. The significant unique functional properties of chitosan, along with its ability to form edible films that can inhibit the spoilage of food, to be used as an active packaging material with a high level of antimicrobial activity or to be used in conjunction with other bioactive materials to serve as a functional surface for the package make chitosan a highly versatile and valuable packaging material [100].

In terms of mechanical properties (i.e., strength), PLA outperforms chitosan and is very compatible with the containers used in traditional food packaging (i.e., cups, trays, bottles). The high crystallinity and molecular weight of PLA also significantly influence its physical characteristics, placing PLA between tough biobased materials and traditional thermoplastics when evaluated from a mechanical standpoint. Therefore, compared with chitosan, PLA can be produced and used at a greater scale; however, the larger scale of use has no effect on environmental protection compared to the use of chitosan. That is, more types of containers can utilize PLA than chitosan; therefore, fewer additives are needed to achieve baseline mechanical properties. Conversely, chitosan may require blends, cross-linking, or additional ingredients to achieve comparable barrier or strength properties, thus resulting in increased production complexity and reduced sustainability post use [173, 176].

Compared with other packaging polymers, PET remains the most mature. Its excellent liquid and gas barrier characteristics, combined with low weight, durability, and ease of processing, have made it the preferred material for a wide range of applications, including bottle, film, and food-contact applications. While PET does not have the same built-in antimicrobial or antioxidant properties as chitosan does, it compensates for this through the reliability, manufacturability, and performance consistency of PET across the wide variety of applications in which it is used. The key difference between PET and other same-sized packaging polymers is that PET has not been developed in an intrinsically sustainable way; however, PET functions efficiently and has been optimized in existing packaging systems. This point is critical because materials that require a large amount of modification to achieve their baseline utility lose much of their relative environmental advantage before they are even used [143, 177].

While glass has great inertness and barrier performance, unlike chitosan, glass does not need to depend on the use of additives or polymer compatibility to achieve long-term chemical stability when it is used in food contact applications. These two characteristics, chemical inertness and long-term food preservation, give glass a significant advantage for food quality preservation. However, both benefits also come at the expense of being brittle, having high mass, and placing transportation burdens on the supply chain for glass. While glass can outperform chitosan in terms of inert barrier performance, it does so at a very different life cycle burden than that of chitosan can provide general performance. Therefore, on the basis of this balanced critique, chitosan should be evaluated primarily for its ability to provide high-value performance in targeted applications without the need for many substantial modifications such that its production burden is disproportionate to the value of the delivered packaging product [148, 178].

### *5.5. Use-phase benefits and system-level constraints*

The use phase of chitosan-based packaging presents very strong arguments for its suitability within food systems where spoilage can occur. Owing to the ability of chitosan to inhibit or stop the growth of microbes and, in selected formulations, enhance barrier performance or have antioxidant effects, it has the potential to extend the shelf life of a product, thereby reducing food loss. This is significant, given that numerous comparative LCAs on food packaging indicate that the environmental impact of spoiled food may be quite large and sometimes even larger than the environmental impact from the packaging material. Therefore, under these circumstances, if a packaging material can effectively reduce spoilage, then the indirect environmental benefits from reduced food waste may be sufficient to offset the increased burden of production of the packaging [168].

The major difference between these two different methods is that one way might have an advantage over another way only if extending the shelf life directly contributes to reducing food waste or to increasing the amount of that food being used as packaging compared with how much would normally be sold in bulk packs. The above statement is essential when looking at chitosan, since the majority of its comparative-based claims are related more to active preservation than to having a low environmental footprint during production processes using chemical-based production. If a particular chitosan material is not able to reduce spoilage while also minimizing its environmental footprint through chemical production and film production techniques, then it could end up being less eco-friendly than its proponents claim it to be.

While PLA, PET, and glass may also provide food preservation benefits, they do so through different mechanisms. The general use of PET and glass is a very strong way to create a barrier and containment for a product, whereas when properly formulated, PLA can provide an acceptable level of performance in some food packaging applications. It is important to note that a benefit during the use phase of a product cannot be attributed to just one material category; the key question in making comparisons between the various types of benefits associated with food protection is the ratio of food protection benefit vs. both the upstream and downstream burdens to provide that benefit. Chitosan may provide significant benefits, but its use-phase benefit needs to be confirmed from a system perspective, as opposed to only through its antimicrobial properties.

### *5.6. End-of-life realism and infrastructure dependence*

Public sustainability messages often become the most misleading at the end of life. While chitosan is marketed, at least in part, as it is a biodegradable polymer, it does not equate to environmentally sound behavior at the time of disposal, as previously discussed. The degradation properties of chitosan depend on many factors, including formula, molecular structure, additive and material selection, and environmental conditions. If chitosan is disposed of with persistent additives or residual solvents or blended with materials that do not degrade, the biodegradation becomes much less valid. In this context, chitosan has a potential end-of-life advantage; however, this is only to the extent that proper formula matching and disposal methods are used [179].

PLA has comparable limitations. While PLA is commonly labeled as biodegradable, the true benefit of PLA at end-of-life relies on the availability of appropriate infrastructure for waste management and proper separation from other types of waste streams. Life cycle assessment (LCA) performed to compare PET and PLA bottles reveals the extreme sensitivity of the apparent environmental benefit of PLA derived from the use of a specific disposal scenario, which indicates that the presumed composting capabilities of PLA, in theory, do not necessarily equate to its performance being better than that of PET. This comparison highlights that both PLA and chitosan suffer from the same overarching issue: a material with an environmentally friendly label does not equate to a system that is environmentally friendly.

Unlike materials made from polylactic acid (PLA), which is made from renewable resources and is biodegradable, polyethylene terephthalate (PET) is not readily biodegradable and is environmentally persistent under normal conditions. However, PET does have well-established recycling programs in place; therefore, the end-of-life issues in the case of PET are complex for several reasons. Recycling mixed materials will also result in lower recycling rates for both materials than for single-component materials. Subsequent research, which includes the study of PET and PLA polymer blends and the study of PET/chitosan polymer blends, demonstrates that simply adding renewable or biodegradable materials in combination with existing PET systems does not necessarily improve the degradability or recyclability of these components; therefore, both studies show that the addition of renewable or biodegradable components complicates the recovery and quality of the recycled product unless the end-of-life method is specifically designed for composite materials. However, they demonstrate how adding renewable or biodegradable materials will lead to compatibility issues unless these materials are designed for use within existing systems [180].

Although many view glass as having very good opportunities for recycling because of how easily it can be made into new products and because it has circular characteristics compared with other materials, the environmental impact associated with glass may vary widely because of factors such as collection and contamination control, transportation distance, the percentage of cullets used, and the energy consumed during melting in production processes. If a large amount of recycled content exists in the glass being melted, the manufacturer will have much greater environmental benefits; however, since glass recovery systems are often not efficient and since there is so much energy used in melting and transporting glass, there is a great deal of burden still associated with melting and transporting glass. Thus, even a material that would otherwise be viewed with the most inherent circularity is reliant on its infrastructure to assist in determining how it will ultimately be recycled. Overall, the conclusion is that a material's end-of-life sustainability is not just an attribute of that material but rather a characteristic of the entire system of which that material is part when it reaches the end of life.

### *5.7. Critical synthesis of comparative trade-offs*

In general, when comparing the various forms of chitosan (a biopolymer), it might not be appropriate to claim that compared with other materials (such as PET, PLA, or glass), chitosan is a superior packaging material, as all of these materials transfer the environmental burden from one place in the life cycle to another. Chitosan largely contributes to the environmental burden in terms of extraction chemistry, washing, drying, and dependence upon additives and the need for realistic disposal (landfilling or incineration); the majority of the environmental burden of PLA lies between the source of the agricultural feedstock, fermentation, and polymerization and their reliance on either composting or favorable disposal scenarios; the majority of the environmental impact of PET lies in their use of fossil feedstock and their long-term durability, but the high efficiency of manufacturing PET and their extensive existing recovery systems is in favor of PET, while the majority of the environmental impact of glass products is contained within the energy used to manufacture the glass products; glass products can provide low environmental impact benefits because of the low use of carbon energy during production [12].

Consequently, the most valuable conclusion drawn from this comparison of products is not that one material “wins” but that claims of sustainability made only on the basis of either feedstock origin or end-of-life labeling cannot be accepted with full reliability. While chitosan has considerable potential because of its unique active functionality and ability to support the reduction of longer-term system burdens (such as food waste) in applications that do not have extreme structural performance, its functional potential diminishes dramatically when an extreme amount of purification, an excessive amount of water/solvent consumption, an excessive number of additives, or uncertain disposal pathways are needed merely to approximate the performance of currently optimized alternatives. Therefore, a critical review must differentiate between the functional promise of chitosan and existing process limitations associated with the sustainability of obtaining its function.

Therefore Chitosan shifts burdens toward extraction chemistry, washing, drying, additive use, and uncertain disposal. PLA shifts burdens toward agricultural feedstock, fermentation, polymerization, and composting assumptions. PET shifts burdens toward fossil feedstocks and persistence, although it benefits from efficient manufacturing and established recycling in some regions. Glass shifts burdens toward melting energy and transport mass, but recycled cullet and low-carbon energy can reduce impacts.

All of these materials are used across a wide variety of applications within the food packaging industry, and each offers distinct advantages and disadvantages in terms of cost, environmental impact, and product quality. **Table 5** discusses the competing alternative materials for food packaging along with chitosan.

**Table 5.** Comparative Analysis of Competing Materials

| Material | | Utility in Food Packaging | Mechanical Properties | Chemical Properties | Physical Properties | Thermal Properties | Barrier Properties | Processing properties/manufacturing | Economic Data | Recycling/End-of-Life Options | Biodegradability | Effect on CO2 emissions/LCA environmental impact | Food-contact safety/regulatory status |
|---|---|---|---|---|---|---|---|---|---|---|---|---|---|
| Chitosan | Polysaccharides | (Cellulose, alginate, starch) utilized for advanced food packaging with improved properties | - Increased tensile strength (cellulose additive has most significant enhancement)<br>- enhanced ductility [181] | -hydrophilic<br>- more compact network structures<br>- chitosan matrix interacts with anionic polysaccharides [182] | - some polysaccharides increase film thickness<br>- increased density<br>- typically transparent or translucent [183] | - improves thermal stability of material [181] | - improves oxygen permeability<br>- increased water resistance [12] | Polysaccharide films are commonly processed by solvent casting, extrusion, and thermomechanical mixing; extrusion is often more scalable for industrial production, while chemical processing can alter polymer structure [184] | Polysaccharides are generally considered low-cost, widely available biopolymer feedstocks, but purification, drying, and film-processing steps can raise total manufacturing cost [185] | Polysaccharide-based films are typically suited to composting or organic recycling rather than conventional plastic recycling, because they are biodegradable and can break down in soil or composting environments [186] | - Biodegradable<br>- readily degrades in soil and within environmental conditions | Using polysaccharides as biobased alternatives can reduce reliance on fossil plastics, but life-cycle results depend on feedstock production, processing energy, and end-of-life route; composting and recycling choices strongly affect total emissions. [187] | Polysaccharides such as starch, cellulose, alginate, pectin, and chitosan are widely studied for food packaging, but food-contact use still depends on formulation additives, migration behavior, and compliance with local regulations. [14] |
| | Plasticizers | - increased flexibility of film-based food packaging | - enhances the film’s elongation to break<br>- decreases tensile strength<br>- reduce brittleness [181] | - reduce intermolecular forces between chitosan chains<br>- alter intermolecular hydrogen bonding [188] | - generally reduces density<br>- smooth, continuous, transparent [189] | - decreases the glass transition temperature<br>- decreases thermal stability of chitosan matrix [26] | - decreased water resistance<br>- increase in oxygen permeability (heavily factorable based on process variables) [181] | Plasticizers lower processing temperature, improve chain mobility, and make polysaccharide films easier to cast, extrude, and thermomechanically process [190] | Common plasticizers such as glycerol and sorbitol are inexpensive and widely used, but cost increases with specialty or food-grade plasticizers and with the need to control migration [190, 191] | Plasticizers can increase flexibility but may also influence degradation and migration during disposal; higher plasticizer content can slow degradation in some systems [192] | - Higher concentration of plastisicer additive slows degradation rates<br>- biodegradable [193] | Plasticizers can improve packaging performance and reduce food waste, but they may add environmental burden through raw-material production and can affect emissions if they increase migration or reduce recyclability [192] | Plasticizers for food packaging must be food-contact compliant and migration-limited; not all plasticizers are suitable for food applications [192] |
| | Crosslinking Agents | Improve water resistance, barrier performance, and shelf-life-related stability of biodegradable films [85] | Can increase tensile strength and rigidity, but too much crosslinker may reduce flexibility and elongation | Create stronger interactions between polymer chains, often through esterification or hydrogen bonding, | Produce denser, more compact films with lower solubility and swelling [85] | Usually, improve thermal stability and delay moisture loss/degradation [85] | Reduce water vapor permeability by tightening the polymer network [85] | Performance depends on crosslinker concentration; excess crosslinker can make films brittle or act like a plasticizer [85] | Usually, low-to-moderate cost, but the total cost depends on the agent, dosage, and processin | Crosslinking can make films harder to reprocess or compost if the network is too dense, so end-of-life performance depends on | Stronger crosslinking can slow water uptake and sometimes reduce biodegradation, but biobased films can still remain | Crosslinking can reduce food spoilage and packaging waste, which may lower life-cycle impacts, but stronger chemical | Safety and approval depend on the specific crosslinker used; food-contact use requires regulatory review |
| | | | [85] | which reduces free hydrophilic groups [85] | | | | | g step [194] | the crosslinker and degree of crosslinking [194] | biodegradable [195] | crosslinking can also increase processing burden if it uses more additives or energy [195] | [195] |
| Polyethylene Terephthalate (PET) | | Commonly used in beverage packaging and sheets/films | - Tensile strength can range from 55-75 MPa with a higher crystallinity corresponding to higher value<br>- Elongation at break percentage range from 65% up to 320% with higher crystallinity providing a lower value [196] | - Chemically inert to most solvents<br>- Can breakdown under contact with strong acids and bases [172] | - Quench cooling during production can result in an amorphous structure<br>- Thermal or stress induced crystallization results in a semicrystalline structure [196] | -It can be used in temperatures up to 55-65°C in its amorphous form and 120°C with a semicrystalline structure<br>- It can also maintain impact resistance at temperatures down to - 40°C [172] | - Has low oxygen permeability and good water vapor barrier<br>- 1.5 $cm^3$ mm/$m^2$ day atm oxygen permeability coefficient for amorphous PET [172] | - Formed through the copolymerization of two crude-oil refining derivatives, ethylene glycol and terephthalol acid [197] | - Production processes are energy-intensive [198]<br>- Recycling of PET is considered low cost [172]<br>- PET bottles can have a recovery rate of up to 93% [197] | - Material recycling results in thermal degradation of the polymer matrix [199]<br>- During the copolymerization step in PET processing, emissions produced can be high [198] | - Resistance to both hydrolytic and enzymic degradation making it nonbiodegradable naturally [199] | - PET manufacture is linked to high fossil energy consumption and greenhouse gas emissions<br>- Material recycling contributes to $CO_2$ emissions [199] | - FDA approved for food contact with specifications depending on structure and form of PET [200] |
| Polylactic Acid (PLA) | | Valued for its transparency and rigidness for cups, boxes | - High tensile strength<br>- Flexural | - Highly sensitive to moisture (especially | - Transparent<br>- Ranges | - low glass transition temperature (approxima | - Highly hydrophobic; however, | - Low heat deflection temperature<br>- | Higher production costs due to | - Mechanical recycling through physical | Biodegradable material derived from corn | - Lower greenhouse gas emissions | - Nontoxic and food safe<br>- |

| | | | | | | | | | | | | |
|---|---|---|---|---|---|---|---|---|---|---|---|---|
| | and containers | strength up to 140 MPa<br>- Elastic modulus (stiffness) 5-10 GPa<br>- Low flexibility Properties are dependent on processing conditions [201] | at elevated temperatures)<br>- Antibacterial [202] | from semicrystalline to amorphous<br>- more dense material at 1.21-1.25 g/cm$^3$ [203] | tely 60°C)<br>- Low melting point (approximately 175°C) [203] | permeable to water<br>- oil and water resistant | Polymerized by converting lactic acid into lactide then using ring-opening polymerization to create PLA<br>- Resulting resin from polymerization is either processed through injection molding, casting, or extruded [202] | reliance on agricultural feedstocks, specialized equipment, and requires high energy input. [203] | reprocessing<br>- Chemical recycling through chemical depolymerization<br>- Solvent-based Recycling<br>- Landfill<br>- Industrial Composting<br>- Does not degrade easily without industrial conditions [205] | starch and other natural sugars (agricultural wastes) [206] | compared to conventional plastics (emits 50%-70% less $CO_2$)<br>- Diminishes plastic pollution, as it breaks down naturally [207] | Approved by FDA for food-contact (lower temperatures) [208] |
| Glass | - Inorganic silica based material<br>- excellent ability to preserve product quality therefore long shelf-life [142] | - High compressive strength<br>- brittle<br>- low tensile strength<br>- high Young's Modulus [142] | - Chemically inert material with virtually all products [142] | - amorphous solid<br>- density at 2.4-2.6 g/cm$^3$ [143] | - low thermal conductivity (high thermal resistance)<br>- glass's coefficient of linear expansion is 9 x 10-6 m/mk [143] | - impermeable to gases and vapors<br>- excellent gas and moisture barriers [142] | - processed through batching and melting (with added recycled glass material) [142] | - high production and transport costs | - 100% recyclable material [142] | - nonbiodegradable, but infinitely recyclable [142] | - high production energy<br>- reuse lowers carbon footprint [142] | - inert and FDA approved for food contact [142] |

***Note:** It is important to mention that the summary given in this table applies to the specific materials and study mentioned. Certain characteristics such as thickness, structure of packaging, formulation, conditions for testing and food applications will not allow us to compare the items chosen for analysis directly. Environmental benefits cannot be obtained by looking at one specific material unless one can show relevant performance of the packaging throughout its life cycle.*

### *5.8. Dominant Environmental Contributions and Cross-Stage Trade-off Dynamics*

Across the comparative literature, upstream extraction and purification emerge as the most consistent environmental hotspots in chitosan-based packaging systems. Although chitosan benefits from being derived from crustacean by-products, this advantage does not automatically translate into low-impact packaging because the conversion of shell waste into purified, food-contact-grade chitosan commonly requires acid–alkali treatment, repeated washing, and thermal input. Industrial LCA evidence indicates that these burdens can be driven either by chemical production or by energy use, depending on supply-chain geography, plant configuration, and the targeted chitosan grade [18]. Accordingly, the environmental profile of chitosan is governed less by feedstock origin alone than by the intensity of the extraction and purification system.

Manufacturing and formulation constitute the second major control point. Pure chitosan films often require plasticizers, blending agents, crosslinkers, or inorganic nanofillers to reach acceptable mechanical, barrier, or antimicrobial performance. These additions can substantially improve functionality, but they also change the interpretation of sustainability. Plasticizers and compatible biopolymer blends may support film flexibility and processability, whereas inorganic nanoparticles and crosslinking strategies can introduce migration concerns, alter degradation behaviour, and leave persistent residues after matrix breakdown. Thus, the gain achieved during the use phase may be accompanied by new uncertainty at end-of-life or during food-contact exposure.

The use phase is the principal domain in which chitosan can offset upstream burdens, but only under specific conditions. Where antimicrobial or barrier performance measurably extends shelf life or reduces spoilage of high-impact foods, the avoided burden of food waste may outweigh part of the material and processing burden. This offset is strongest when packaging is targeted to perishable products with demonstrated preservation benefit rather than applied as a generic substitution. By contrast, if performance gains are marginal, if additional layers or additives substantially increase material throughput, or if consumer convenience effects increase single-use consumption, the net sustainability benefit becomes weak or uncertain.

End-of-life outcomes are best interpreted as conditional rather than intrinsic. Chitosan itself may degrade relatively rapidly in some controlled composting or soil environments, yet actual disposal conditions vary widely across landfill, mixed-waste, wastewater, and aquatic pathways. The presence of additives, crosslinked domains, or residual inorganic components can further decouple biodegradability claims from real environmental fate. Overall, the sustainability of chitosan-based food packaging is therefore determined by a hierarchy of variables: extraction intensity first, formulation strategy second, demonstrated use-phase food-waste reduction third, and disposal infrastructure fourth. A credible sustainability claim should only be made when those variables are evaluated together rather than independently.

## 6. Current Drawbacks and Challenges

While often promoted as a sustainable alternative to fossil fuel-based food packaging, several environmental and commercial constraints related to the viability of chitosan-based packaging exist. These drawbacks are present throughout the entire life cycle of chitosan-based packaging and include feedstock sourcing, chitin extraction, chitosan processing, film manufacturing, additive incorporation, use-phase performance, food contact safety, and end-of-life treatment. Therefore, the sustainability of chitosan-based packaging cannot be inferred solely on the basis of its biobased origin, antimicrobial properties, or ability to biodegrade. Instead, an assessment of sustainability must consider whether the environmental benefits of waste valorization, food preservation, and compostability outweigh the environmental burdens associated with chemical processes, energy requirements, water use, regulatory testing, and disposal uncertainty [209].

Nano-enabled chitosan packaging systems can improve mechanical, barrier, antimicrobial, and active-packaging performance, but the incorporation of nanomaterials may also introduce additional uncertainties related to migration, environmental persistence, ecotoxicity, and end-of-life behaviour. The use of nanofillers may improve the mechanical performance, oxygen barrier performance, antimicrobial performance, and shelf life, but these additional material-specific properties may complicate biodegradation processes and potentially increase the potential for migration or persistence of residues from the chitosan matrix following decomposition. Furthermore, while high-purity chitosan can be obtained using an acid–alkali extraction process for food-contact applications, this extraction process creates significant quantities of chemical effluents and has significant upstream environmental impacts (e.g., water consumption, energy consumption, and greenhouse gas emissions). The current question is whether or not chitosan can be converted into functional packaging; however, the real question is whether or not functional packaging can be produced, used, and composted in ways that demonstrate its net benefits to global sustainability on the basis of its use as a practical packaging solution [14].

### *6.1. Chemical-intensive processing and scale-up burdens*

Using chitosan to package food and other products may be an excellent way to help the environment, but chitosan-based packaging comes with environmental costs associated with the production of chitosan from crustacean shells (waste) and the extraction of chitin from crustacean shells. Crustacean shells can be valorized as a chitin/chitosan feedstock, but they require extensive purification before food-contact packaging use. These shells must undergo various stages of purification (i.e., demineralization, deproteinization, decolorization, and deacetylation) before they can be utilized as food packaging. Chitosan is produced through a multistep process involving demineralization, deproteinization, decolorization, and deacetylation (with strong acids such as hydrochloric acid and strong bases such as sodium hydroxide), as well as repeated washing, neutralization, filtering, and drying to yield chitosan. Therefore, the burden may shift from polymer feedstock extraction toward chemical, water, wastewater, and energy demands used in the production process, producing diluted wastewater, producing salt-containing effluent, and generating the energy to run the production processes. Various life cycle assessment studies have evaluated the effects of the use of raw materials for chitosan production, the use of chemicals only for chitosan production, the discharge of fresh water into the environment, the consumption of electricity for chitosan production, and the disposal of wastewater produced during the production of chitosan, all of which vary the final environmental profile of the supply of chitosan [18].

This creates a paradox of sustainability in terms of chitosan. Chitosan is often marketed as a circular economy material since it can be made from seafood processing waste. However, the transformation of shell waste into high-quality purified chitosan typically requires significant resources. If shell waste is being used as the feed stock but the process relies heavily on reagent concentrations, high temperatures, washing periods and fossil fuel-based energy, then the environmental benefits of using waste feed stock may be lost. This is a critical issue when considering food-packaging applications where regulations and consumer expectations require significant purity of the material with low content in terms of protein, ash and chemical residue. When supplying chitosan products for food packaging applications, food-contact packaging generally requires stricter purification than lower-grade applications such as wastewater treatment or agriculture [121].

There are many alternative methods to extract chitosan, including microbial fermentation, enzymatic deproteinization, organic acids and deep eutectic solvents, and other forms of green chemistry, which have been proposed as ways to reduce chemical intensity in the production of chitosan; however, not all of these methods are yet used in the commercial production of chitosan. Therefore, the use of alternative solvents can present significant challenges when large-scale production of chitosan is considered by replacing hydrochloric acid or sodium hydroxide with a green alternative; thus, an integrated approach is needed to develop a method for producing chitosan with a low overall environmental impact while maintaining yield, producing a high-quality product and being economically viable [12].

### *6.2. Feedstock variability, quality control, and standardization gaps*

Another significant limitation of chitosan is its inherent variability. Chitosan is a heterogeneous biopolymer whose physical characteristics are defined by its origin, shell chemistry, animal species, collection time, storage temperature, storage duration, moisture content, and microbial spoilage, pretreatment method, extraction method, deacetylation conditions, and drying history. The key parameters that affect the production of films, antimicrobial effectiveness, oxygen barrier properties, water vapor permeability, tensile strength, elongation, and biodegradability are the degree of deacetylation, molecular weight, viscosity, crystallinity, residual ash, residual protein, moisture content, and solubility [14].

The main challenge to the repeatability and commercialization of chitosan is its variable nature. Although commercially specified chitosan grades are available, differences in characterization methods and incomplete reporting of molecular weight, degree of deacetylation, viscosity, residual ash, protein content, and moisture limit reproducibility and cross-study comparability. In addition, the variety of ways to characterize chitosan, such as "low," "medium," or "high" molecular weight, as well as the lack of mention of the deacetylation level, viscosity, ash content, and protein residual level, allows the manufacturer to use data from the testing laboratory; however, when the marketer/user begins to develop using the products that are in production, the manufacturer does not adhere to the same level of accuracy in the specific area of manufacturing their packaging formulations in terms of thickness, sealability, durability, migratory nature, and physical durability.

The differences in manufacturing standardization do not allow for reasonable comparisons across different chitosan films, as chitosan materials can be collectively referred to as "biodegradable packaging from chitosan."For example, two chitosan films may differ substantially in molecular weight, deacetylation degree, and residual ash content and have inherently higher molecular weight with higher rates of deacetylation measured in comparison to the other chitosan film's lower molecular weight and residual ash. The individual properties of chitosan affect its antimicrobial activity during food media interactions. Additionally, other external factors affect the antimicrobial ability of chitosan in the environment. There can be great variability in physical activity and how individual chitosans are made; therefore, the reason for apparent improvements in antimicrobial activity will not be known until the reason for improvements is identified because of the creation process used for producing the films and the environmental conditions during testing [132].

This variability in the production of chitosans has further implications for life-cycle assessment. Studies performed using chitosan produced by different methods (e.g., through different extraction methods, with different levels of chemical input, and with differing levels of chemical purification) may present varying environmental impacts. For example, one chitosan film that has been manufactured from the shells of shrimps and has undergone optimized chemical recovery will have a much lower environmental impact than one that has been manufactured from shrimp sourced from a long distance through a low-yield extraction procedure and without chemical recovery. Therefore, standardizing the feedstock and process used to create chitosan is not only a concern of material science; such standardization is required to develop credible sustainability assessments. Consequently, life-cycle inventory data obtained for one chitosan grade or extraction route should not be transferred directly to another without considering differences in processing yield, purity requirements, chemical consumption, energy demand, and functional packaging performance.

### *6.3. Functional performance limitations under real food-packaging conditions*

The advantages of using chitosan as a food packaging material include the ability to form films, act as a barrier to oxygen, inhibit microbial growth and be an antioxidant, and be suitable for use with active ingredients. While neat chitosan films have a number of advantages for use in commercial packaging, they do not always meet all functional packaging requirements. The major disadvantage of neat chitosan films is moisture sensitivity; the hydrophilic properties of chitosan tend to lead to lower water–vapor barrier performance than those of hydrophobic polymers from petroleum sources, such as polyethylene and polypropylene. Increased humidity can cause the chitosan film to swell, decrease its mechanical strength, increase its plasticization, and decrease its barrier performance. Therefore, the applicability of neat chitosan films in the packaging of high-moisture foods, refrigerated products, seafood, fresh meats, and produce with high levels of respiration or water activity is limited [132].

Another issue with chitosan films is that they can be brittle and exhibit low elongation at break, particularly under dry conditions or when not enough plasticizer is used. Such brittleness can create difficulties when these films are handled, folded, sealed, transported and stored. Commercial food packaging also needs to withstand mechanical stress during filling, distribution, consumer handling and shelf display. If a film cracks, loses its integrity or fails at the points where it has been sealed, then its environmental benefits will not be achieved because the amount of food that spoils as a result of poor packaging will be greater than the savings achieved through reduced use of packaging. This is why chitosan films are frequently combined with other types of polymeric materials, plasticizers, nanocellulose, starch, PVA, proteins, clays, essential oils and metal/metal oxide nanoparticles [12].

When polymers are modified, some compromises are made as a result of the modifications. For example, plasticizers (such as glycerol and sorbitol) improve the flexibility of a polymer but can also lead to increased moisture absorption and lower tensile strength if the plasticizer content is too high. Crosslinking increases water resistance and creates stronger bonds in the material but might reduce the biodegradability of the material or create regulatory challenges related to the type of crosslinked polymer produced. Nanofillers can increase the mechanical performance of the polymer, oxygen barrier characteristics, UV (ultraviolet) protection, and antimicrobial performance; however, whether these benefits are actually functional depends upon the quality of the dispersion of the nanofiller, the amount of nanofiller used, the compatibility of the nanofiller with the polymer matrix, and the long-term stability of the nanofiller in the polymer matrix. When a greater amount of nanofiller is used, clumping can occur, some optical clarity can be lost, and they may complicate compostability or leave persistent residues, depending on additive chemistry and loading of the polymer [121].

Although a material may exhibit superior performance under laboratory conditions, it may still not perform well in actual commercial food packaging environments. Studies conducted on chitosan films often utilize a simplified testing environment (e.g., constant humidity, short contact time, model microorganisms, or isolated mechanical tests) as opposed to the interactions that occur under actual conditions (e.g., food chemistry, temperature, humidity, pH, fat content, oxygen exposure, microbe load, handling stress and shelf life). As such, future film evaluations need to transition away from the isolated optimization of film properties toward the testing of chitosan films under realistic food-contact and supply-chain conditions [77].

### *6.4. Nanomaterial safety, migration, and regulatory uncertainty*

Chitosan packaging can be enhanced with inorganic nanofillers, including clays, silica, metal oxides, and metallic nanoparticles; however, nano-enabled food-contact materials require formulation-specific migration and toxicity assessment. The addition of metal and metal oxide nanoparticles (e.g., silver, zinc oxide, titanium dioxide, copper oxide, silica, and clay) may provide antimicrobial, UV blocking, mechanical and barrier properties. Therefore, these nanomaterials should be evaluated for their use in food-contact packaging as a result of their potentially different behavior in food compared with their respective bulk materials. Migration or release of nanoscale constituents may occur under particular food-contact conditions, but the extent and form of release must be established for the specific formulation and exposure scenario. The various factors that affect the migration of nanoscale material into food include particle size, surface chemistry of the particles, concentration of the particles, aggregation state of the particles, polymer–matrix interactions, food composition, pH of the food, fat content of the food, temperature, length of time stored, and mechanical shearing [210].

Migration is critical because food-contact packaging may release substances into food during storage, handling, or disposal. While nanoparticles may be encapsulated in chitosan matrices, they still have the potential for migration via diffusion, swelling, degradation of the matrix, wear and tear, and through incidental contact with foodstuffs with a pH <7 or fatty nature.

The migration may involve intact nanoparticles or ionic dissolved or transformed particles. For example, while there is a positive potential for the use of silver nanoparticles with respect to antimicrobial activity, there is potential for concern associated with the migration of silver ions, cytotoxicity, antibiotic resistance development, and the accumulation of silver in the environment. Similarly, ZnO and $TiO_2$ nanoparticles may provide antimicrobial or UV-protective functions, but their safety depends on particle size, solubility, exposure route, and transformation during use and disposal.

Regulatory frameworks increasingly acknowledge the need for adequate and specific assessments of nanomaterials. EFSA guidance for food indicates that the presence of small particles (including NPs) must be demonstrated and that the physicochemical characteristics of the small particles and their exposure must be assessed. Similarly, FDA guidance indicates that the use of nanotechnology in a product may need to be evaluated more closely than that in non-nano products because dimension-based characteristics potentially impact the product's safety, effectiveness, quality, or regulatory status [211].

Regulatory compliance must be evaluated for the complete food-contact formulation and its intended conditions of use rather than by applying a single migration threshold to all chitosan-based materials. In the European Union, food-contact materials are subject to the general safety requirements of Regulation (EC) No 1935/2004, while Regulation (EU) No 10/2011 establishes specific requirements for plastic materials and articles, including substance-specific migration limits (SMLs) and an overall migration limit of 10 mg/dm$^2$ for applicable plastic food-contact materials. Substances in nanoform may only be used within the scope of this Regulation when explicitly authorized and specified for that form. In the United States, constituents of food-contact materials that are reasonably expected to migrate must be covered by an applicable authorization, such as provisions within Title 21 CFR, an effective Food Contact Notification, GRAS status, a prior sanction, or an applicable Threshold of Regulation exemption. FDA evaluation of food-contact substances considers the intended conditions of use, migration into food, resulting dietary exposure, and toxicological information. Consequently, measured migration of residual acetic acid, plasticizers, chitosan-derived oligomers, and metal or metal-oxide constituents must be interpreted against the substance-specific regulatory status and restrictions applicable to the particular formulation and intended use. Where no formulation-specific migration data or directly applicable constituent-specific limit is available, this should be treated as a regulatory evidence gap rather than as evidence of safety [222, 223].

The challenge for developing food contact packaging based on chitosan nanocomposites arises from the fact that the speed with which new materials are developed exceeds the speed at which regulatory bodies develop regulations to manage these materials. Even if a material is labeled “biodegradable” or “biobased,” new formulations using engineered nanoparticles, antimicrobial agents or active release systems may require additional testing to determine migration and toxicological safety and approval for food contact. Additionally, the possible consumer perception of chitosan as “natural” and/or “biodegradable" and/or “nanoenabled” may lead to confusion in their perception of the safety of these products. Therefore, safety-by-design should be the primary consideration throughout the entire formulation development process, rather than only at the end of the development process, to ensure validation before finalization [212].

### *6.5. End-of-life uncertainty and residual additive persistence*

Chitosan-based packaging is biodegradable, but its biodegradability is often overstated. Factors such as molecular weight, degree of deacetylation, crystallinity, film thickness, crosslinking density, moisture, pH, temperature, microbial activity, and method of disposal ultimately determine the rate at which a material made from chitosan will degrade. While chitosan films may degrade under laboratory composting conditions, it will likely take much longer for them to break down in landfills, marine environments, soil, wastewater systems, or low-moisture waste streams. Thus, biodegradability should be considered a condition-based property, not an absolute property [211].

End-of-life assessments should distinguish neat chitosan films, polymer blends, chemically crosslinked formulations, organic-nanofiller composites, and metal/metal-oxide nanocomposites, since additive identity, loading, crosslinking density, and disposal conditions can alter biodegradation, mineralization, and residual-material behaviour. These differences in degradation paths complicate the process of waste management for end-use applications. In many cases, chitosan can biodegrade while inorganic residue remains. In addition, by increasing the crosslink density or decreasing water absorption, some additives may impede the biodegradation process of chitosan, whereas other additives released during biodegradation may impact soil microorganisms, aquatic microorganisms, and/or the effectiveness of wastewater treatment systems. As a result, an evaluation of end-of-life should consider not only whether chitosan has degraded but also the residues remaining (if any) and the ecological nature of these residues [210].

In particular, organic-nanofiller composites should be assessed for degradation of the complete film, whereas metal and metal-oxide nanocomposites additionally require consideration of persistent inorganic residues, particle transformation, and possible ion release. These outcomes should be distinguished from experimentally demonstrated ecotoxicity, which requires formulation-specific exposure and biological-effect measurements. Importantly, the possibility of additive release or persistence should be distinguished from experimentally demonstrated ecotoxicity. Similarly, film disintegration or mass loss does not necessarily establish complete biodegradation or the absence of residual constituents.

The compostability standards explicitly indicate that compostable materials such as ISO 17088 include disintegration, final aerobic biodegradability, no detrimental effects on land-based organisms, and control of components in the compost material. Therefore, a compostable material is not defined by being biobased or being partially biodegradable. To be considered compostable, the material must meet certain performance standards under certain composting conditions and have no residues that are persistent or hazardous [211].

One more challenge is that there is a mismatch between the design of a material and how we manage waste with the materials we design. Although chitosan-based films are compostable in commercial facilities, this benefit will not be realized until consumers can access the correct collection and commercial composting facilities. If the materials are thrown into a landfill, disposed of via incineration, or mixed into a recycling stream, there is the potential for us to lose the benefit of the material being environmentally superior. Another problem is the potential for consumers not to identify a chitosan film as different from a traditional plastic film, increasing the likelihood of contamination in either a recycling or composting stream. To ensure that the material is sustainably deployed, clear labeling, disposal instructions, and compatible infrastructure must be in place.

### *6.6. LCA data gaps, functional-unit sensitivity, and market adoption barriers*

Finally, the absence of comprehensive life cycle data for chitosan-based packaging systems presents another challenge. Most published studies on chitosan-based packaging systems address only characteristics such as strength, antimicrobial activity, oxygen and water permeability, antioxidant activity, or biodegradation; however, few studies report full lifecycle inventories, including collection of shell waste, transportation of materials, storage of materials, production of reagents, energy usage, water usage, wastewater treatment, solvent recovery, drying the polymers to form films, converting films into packaging, additives to improve the product, losses on the production line due to poor quality control, use phase food waste reduction, and realistic options for post use disposal [18].

The choice of functional unit heavily influences the outcomes of the study. For example, simply comparing 1 kg of chitosan film to 1 kg of plastic film may lead to misleading results because of the many differences in thickness, strength, barrier properties and durability among the different materials. For the purpose of food packaging, an appropriate functional unit may be defined as having the ability to deliver a specified quantity of food under “acceptable” conditions for a predetermined period of time. This finding is important because packaging can have an indirect positive effect on the environment by lowering food waste, which should be demonstrated through real food systems, which are not assumed to be based on the antimicrobial properties of the packaging [213].

The cost of chitosan packaging and other integration-related issues currently limit the amount of market penetration that chitosan packaging can obtain. Most of today's packaging has been designed to accommodate petroleum-based polymers and is optimized for high-throughput processing. Films of chitosan that have been developed using either a solvent cast or as a coating on a lab scale may not be able to be used on an existing roll to roll, extrusion, lamination, filling or sealing operations. Unless the manufacturers are confident that chitosan packaging can meet the necessary performance standards, regulatory requirements, price expectations, and reliability of supply, chitosan will not be widely accepted by manufacturers. Consequently, broader commercial adoption of chitosan-based packaging will depend on whether manufacturing costs, production capacity, performance requirements, regulatory compliance, and supply-chain reliability can become competitive with established packaging systems. However, the best opportunity for the widespread use of chitosan is in applications where its unique active capabilities (particularly its antimicrobial and antioxidant properties) can provide significant measurable improvements in the shelf life of the product and warrant a price premium compared with those of conventional polymers. These comparisons must take into account the complete packaging structure, including all the supporting layers or active substances used to simulate similar protection, and should differentiate between the food preservation effects obtained in practice and the inferred reduction of food waste. Consequently, future comparative assessments should explicitly report both the material reference flows required to provide equivalent protection and the processes included within each system boundary.

Economic viability is also affected by several factors, including feedstock gathering and transport costs, extraction and purification rates, reagent and solvent reuse, energy consumption, wastewater disposal, drying processes, additives' costs, and manufacturing losses. Therefore, all these factors should be taken into account while performing feasibility studies on the industrial scale. Thus, techno-economic evaluations should supplement ecological ones with the information necessary for assessment of real production potential. Nevertheless, sufficient and directly comparable data on cost-effective production are still rare, which leads to absence of universal economic ranking of chitosan-based packaging solutions.

## 7. Outlook and Recommendations

In future studies, the formulation of chitosan that shows the functional benefits of use in comparison with credible commercial packaging should take the greatest priority. The highest priority applications may include packed food with measurable properties and benefits in terms of antimicrobial activity, oxygen barrier properties, or controlled release of active ingredients. However, such applications should be concerned only if the new properties can be assessed and relatively high levels of extraction and feasibility of producing, migration safety, and final use conditions are ensured. Therefore, future work should not view chitosan as a "one-size-fits-all" alternative to conventional fossil fuel plastics. In addition, compared with other alternatives, chitosan-based packaging materials, depending on their unique combinations of waste-derived materials and the presence of various antimicrobial properties, do not have better environmental performance in every environmental impact category, every type of food, or under every type of disposal condition.

Most of the chitosan-based packaging successes observed to date have occurred within the context of active packaging, edible coatings, oxygen-sensitive products, or highly perishable foods and supply chains that might benefit from extending the shelf life of food products that reduce possible food waste.

### *7.1. Development of pathways for lower-impact and closed-loop chitosan production*

Research aimed at creating demand-focused methods of extraction and conversion with reduced chemical intensity and enhanced product performance should include alternative methods such as enzymatic deproteinization, microbial fermentation, organic acid demineralization, deep eutectic solvent systems, microwave-assisted extraction, ultrasound-assisted extraction and integrated biorefinery approaches. This is in addition to the further development of more traditional methods. When these methods are evaluated for industry relevance, standard laboratory yield or reagent substitution is not sufficient for evaluating industrial performance. Instead, industrial metrics such as total chitin recovery, the degree of deacetylation, the preservation of molecular weight, residual protein, total ash, total water demand, total energy demand, total wastewater load generated, total solvent recovered, total time to process, and total cost will be used to evaluate these alternative extraction/processing methods [18].

A focus on closed-loop chemical management must be a priority for designers. Recovery of acid and alkali, recycling of water, heat integration, drying at low temperatures, and minimization of wastewater reduce both the environmental impact and per tonne of shell waste processed. Producers need to determine how to turn other fractions (non-chitin) of shellfish waste (protein, minerals, and pigment) into value-added products instead of simply considering them as waste. This change will convert chitosan production from an extraction-only model to one in which crustacean shell waste is refined for use in new and innovative ways. This will enhance the material efficiency of chitosan production and provide chitosan-based packaging with a solid foundation in the circular economy [18].

To evaluate whether alternative extraction processes provide true ecological benefits, it is necessary to evaluate them in terms of their chemical use, energy and water demands, solvent reclamation, operational yield, and waste treatment impacts in comparison to normal methods under similar life-cycle assumptions. A process that employs less aggressive reagents but requires longer processing times, has poor solvent recovery/evaporation characteristics, or has high energy consumption for drying may not exhibit an overall reduction in environmental impact. Future research into chitosan production processes should combine material characterization studies and life-cycle inventory (LCI) development to help achieve optimum performance of both the environment and materials concurrently.

### *7.2. Establishing standardized chitosan grades and food contact specifications*

The development of uniform chitosan types for use across food-packaging applications should occur via standardized chitosan grades. Future studies and suppliers should develop consistent methods for documenting where the species that provides the chitosan came from, how the chitosan was derived, such as via which feedstock, the degree of deacetylation, molecular weight range, viscosity, ash (inorganic solid materials that remain after the burning away of organic matter), residual protein, and moisture content, crystallinity and solubility, and contamination levels, particularly through the numbers of microorganisms found in the tested materials (i.e., pathogen spores, toxins or chemicals), as well as residual levels of reagent chemicals. Furthermore, the above parameters should be related to the functional properties of the materials being investigated (e.g., tensile strength, elongation, gas permeability, water vapor permeability, antimicrobial properties, clarity or transparency, sealability, and degradability). Consistent reporting of these parameters, together with the analytical methods used to determine them, would improve the reproducibility of material studies and support more reliable comparisons of life-cycle inventories across production routes and packaging formulations.

Whereas information regarding the identity of additives and their amounts, particle size distribution, surface chemistry, method and extent of dispersion, state and manner of particle aggregation or clumping, manner of migration of the component within the food, and residue characteristics upon disposal must be provided for any published data related to nanocomposite packaging (this information is essential to reproduce results and ensure the safety and sustainability of the material), this information will be especially critical for active packaging systems since the activity of antimicrobial or antioxidant agents present will depend on controlled release, stability of the active agent, and surface interactions [121].

Testing for food-contact uses should be designed on the basis of actual use cases; for example, the testing of a chitosan coating for dry bakery products cannot be treated in the same way as the testing of a chitosan film for seafood, meat, acidic fruit, or high-fat food products. Numerous elements (e.g., temperature, humidity, pH, time of contact, food simulants, fat level, and mechanical handling) affect the results of food-contact testing and are not consistently reported across studies. Having standardized reporting of these elements, however, would increase the ability to compare results between studies and facilitate the approval process by regulatory authorities, as well as help identify the grade of chitosan that is appropriate for use within specific types of food packaging [132].

### *7.3. Advance scalable and low-energy nanocomposite manufacturing*

Scalable manufacturing will be a major focus of process development. While solvent casting is currently still used in laboratory R&D, it has several reasons for not being used in high-speed packaging processes: (i) extended dry times, (ii) solvent safety, (iii) batch variability from sample to sample and (iv) slow processing times. Some of the more industrially applicable techniques are roll-to-roll coating, spray coating, extrusion-compatible blending, lamination, continuous casting, and aqueous coating systems. All of these manufacturing processes need to be enhanced with respect to reducing energy consumption, using minimal solvents, providing consistent film thicknesses, obtaining a stable distribution of additives, and being compatible with current packaging equipment [14].

Results of pilot studies must include specifications related to production rate, yield, drying energy consumption, production cost, equipment need, and variability of the process to evaluate the technical and economic feasibility of production. An integration of this information with life-cycle inventory would facilitate the evaluation of various economic aspects in the transition to the industrial scale of production.

The optimal strategy for designing nanocomposites is an additive-minimization approach, whereby additives are only added when they can provide measurable benefits, such as a high barrier to UV radiation or moisture, high tensile strength, antimicrobial properties, and the oxygen-scavenging and antioxidant activity. Natural biodegradable additives, such as nanocellulose; different forms of starch, proteins, essential oils and phenolic compounds; and biobased plasticizers, should be preferred. Inorganic nanoparticles should be used only if their performance exceeds concerns about their potential migration, toxicity and/or residue left behind [12].

Another approach to designing packaging is multilayer constructions; for example, instead of relying on chitosan to perform all functions as a packaging material, it can be used as an active or barrier material within a larger/whole packaging structure. Chitosan can act as an antimicrobial coating on paper, cellulose, PLA or other biobased substrates. The creation of multilayer packaging systems and how they will work best will depend upon design considerations, with the goal of providing a functional yet easily recyclable or compostable material that maintains compatibility with the intended recycling, composting, or disposal pathway.

### *7.4. Strengthening circular seafood waste valorization and policy support*

The circularity of chitosan packaging depends partly on the sourcing and management of seafood-processing residues. The collection process for shell waste should be organized by region to minimize the transport burden, prevent spoilage, and maintain consistency of quality in both the feedstocks and the finished product. It is also necessary to establish an infrastructure that provides the storage, sorting, washing, and pretreatment of seafood-processing waste, subsequently allowing these wastes to be processed into a reliable raw material for the industrial production of chitosan. Failure to establish such an infrastructure results in supply instability, contamination, and an increased cost of processing for producers of chitosan.

Government and industry associations could support seafood waste valorization by providing grants to develop regional biorefineries, incentivizing seafood-processing waste segregation and collection, creating plans for food waste composting infrastructure, using renewable energy to process food waste, and creating procurement plans for certified low-impact packaging items. Systems for extended producer responsibility could also incentivize the design of convenient packaging for the intended end of life.

Chitosan should not be assumed to be sustainable, and packaging claims for chitosan should only be made with third-party testing and clear descriptions of disposal conditions. We should avoid creating a resource-driven dependency on chitosan. Chitosan should not become a driver of increased crustacean harvests, so any increase in demand for chitosan to create food packaging should prioritize existing seafood-processing waste streams and avoid incentivizing additional crustacean harvesting. Traceability and responsible sourcing can help maintain the value of chitosan in the circular economy [214].

### *7.5. Use application-specific LCAs and responsible deployment*

A primary recommendation is to perform an application-specific life-cycle assessment to evaluate a specific chitosan-packaging application. Future life-cycle assessments (LCAs) should utilize international standards set forth in the International Organization for Standardization (ISO) and include the following: the sources of raw materials, transportation, extraction chemicals, water consumption, energy consumption, wastewater treatment, drying processes, manufacturing of the chitosan film, production of additives, waste generated from the production process at the packaging line, effectiveness of the packaging throughout its use and during the time it is used, the ability of the packaging to reduce food waste, and optimal disposal at the end of the service life of the product.

A functional unit should be representative of the actual function provided by packaging. Rather than comparing the same mass of materials in the LCA assessment, each LCA should evaluate the ability of the packaging to contain a specific amount of food in a reasonable state for a defined shelf life. This evaluation is most relevant for active chitosan packaging because the environmental benefit of using active chitosan would not necessarily be from displacing other materials but rather through decreased food spoilage. Throughout the evaluation process used to create an LCA, food waste reduction resulting from packaging should be emphasized, particularly for food products with the highest level of environmental impact, which include meat, seafood, dairy and produce [213].

This means that the use of chitosan is responsible for the significant functionality or value of its use in the environment. There are several potential uses for chitosan because of its high potential value, including antimicrobial coatings for perishable foods, products that are sensitive to oxygen, edible coatings on fresh produce, active films for highly perishable meat and seafood, and bulk or secondary packaging, where shelf-life benefits can be demonstrated where an extended shelf life can be proven. However, for food items with minimal spoilage or with chitosan requiring significant alteration to comply with basic packaging requirements, more suitable packaging materials may exist. This will lead to future packaging being created with a narrower market focus as opposed to everyone using a single biopolymer-based packaging material.

Based on the analysis of life-cycle trade-offs, it is possible to establish the application-centric decision-making framework which may distinguish among three sustainability outcomes. The environmental benefit can be considered if traceable raw materials, energy-efficient processes and formulations, and evidence of effectiveness are coupled with life-cycle assessment being able to show lower environmental impacts under certain conditions. The case is uncertain if there is a lack of necessary knowledge on ingredient amounts, indicators of functional equivalency, reduction of food waste, safety, and behavior at the end of the life cycle. There is potential for environmental harm if the requirements for chemicals or energy, use of packaging materials, effectiveness of food preservation, or other problems of disposal exceed the benefits. Therefore, the criteria presented provide a framework which helps to evaluate chitosan formulations rather than simply considering their renewable nature or biodegradability.

For practical deployment, chitosan-based packaging should be evaluated as part of a broader sustainable-packaging strategy for packaging materials. Success will depend on how materials are related to overall production, the standardization of material, the safety of additive design, realistic evaluations of material performance, the validation of end-of-life claims, including compostability and biodegradation under specified conditions at the end of their life cycle, and overall life cycle evaluation showing true environmental reductions rather than burden shifting during the manufacturing or disposal process.

## 8. Conclusions

Chitosan-based food-packaging systems are promising because of their potential for waste-derived sourcing, film-forming ability, antimicrobial functionality, and biodegradation under suitable conditions. However, the evidence reviewed in this work demonstrates that none of these characteristics alone establishes environmental superiority. The sustainability of chitosan-based packaging is instead governed by interconnected trade-offs involving feedstock sourcing, extraction and purification, formulation, packaging performance, food-preservation benefits, and end-of-life behaviour. Therefore, chitosan cannot be considered environmentally favourable solely because it is biobased, waste-derived, or biodegradable.

Across the reviewed evidence, upstream extraction and purification emerge as recurring environmental hotspots. Acid–alkali treatment, repeated washing, drying, wastewater treatment, and energy demand can substantially reduce the environmental advantage associated with valorizing crustacean shell waste. The analysis further identifies a hierarchy of important sustainability controls: extraction and purification intensity determine much of the upstream burden; formulation strategy influences both packaging performance and downstream uncertainty; demonstrated food-waste reduction determines whether use-phase benefits can compensate for production burdens; and disposal infrastructure determines whether biodegradability translates into a meaningful end-of-life advantage. Consequently, the environmental performance of chitosan depends much more on how the material is produced, formulated, used, and disposed of than on its biological origin alone.

A further finding is that direct environmental comparison across packaging materials remains constrained by methodological differences in the available literature. Published studies use different functional units, system boundaries, product grades, allocation or substitution assumptions, processing inventories, and end-of-life scenarios. Equal-mass comparisons are particularly limited when materials provide different mechanical, barrier, antimicrobial, and preservation functions. For food-packaging applications, future comparative life-cycle assessments should therefore use functionally equivalent units that reflect the actual packaging service provided, including the quantity and type of food protected, required storage period, packaging performance, and, where experimentally demonstrated, avoided food waste. This is particularly important for active chitosan packaging, where the principal environmental benefit may arise from reducing food spoilage rather than simply displacing an equal mass of conventional packaging.

Formulation also strongly affects the sustainability outcome. Plasticizers, polymer blends, crosslinkers, essential oils, nanofillers, and other functional additives can improve mechanical strength, moisture resistance, barrier performance, antimicrobial activity, and shelf life, but these improvements may also alter migration behaviour, degradation kinetics, recyclability, compostability, and residual-material persistence. End-of-life performance should therefore be evaluated for the complete formulation rather than inferred from the biodegradability of the chitosan matrix alone. Similarly, shelf-life extension should not automatically be interpreted as food-waste reduction; this benefit should be demonstrated under realistic food-storage and distribution conditions before it is credited as an environmental advantage.

Among the quantitative LCA evidence examined, Muñoz et al. [18] demonstrates the strong sensitivity of chitosan-production impacts to processing route, energy supply, feedstock management, by-product treatment, and supply-chain configuration. The substantially different environmental burdens reported for the investigated supply chains should not be interpreted as a direct regional comparison because the systems produced different grades of chitosan for different applications. Rather, they demonstrate how substantially environmental performance can vary within the same nominal material family. Overall, the reviewed evidence does not support chitosan-based packaging as a universally sustainable replacement for conventional materials. Its strongest environmental case exists in targeted applications where efficient production, limited formulation burden, demonstrated food-preservation performance, food-contact safety, and compatible end-of-life infrastructure occur together. Where these conditions are uncertain or absent, the environmental advantage is likewise uncertain, and burden shifting may occur instead of an overall reduction in environmental impact.

The sustainability paradox of chitosan-based food packaging therefore lies in this balance between biological origin, functional performance, processing burden, and realistic end-of-life behaviour. Future development should move away from broad sustainability claims based on material identity and toward application-specific evaluation of complete packaging systems. More consistent life-cycle methodologies, functionally equivalent comparisons, transparent reporting of process inventories and allocation assumptions, formulation-specific safety and degradation studies, and experimentally demonstrated food-preservation benefits will be necessary to establish when chitosan-based packaging represents a genuine environmental improvement.

---

## Declarations

## Competing Interests

Authors declare no competing interests.

## Data Availability

No new datasets were generated or analyzed during this study. All information supporting the findings of this review is available within the manuscript and its cited references.

## Funding

This work was financially supported by Vellore Institute of Technology, Vellore, under the Faculty Seed Grant (Sanction Order No. SG20210277).

## Ethics, Consent to Participate, and Consent to Publish declarations

Not Applicable.

## Declaration of Ethical GenAI Usage

During the preparation of this work the authors used Grammarly AI and ChatGPT to improve the readability and language of the manuscript. After using this tool/service, the authors reviewed and edited the content as needed and take full responsibility for the content of the published article.

## Author Contribution Statement

Conceptualization, M.C., Mi.C., A.K., S.P., G.W., R.Z., T.D., S.G., R.M.S., R.S., S.L., and U.C.; methodology, S.P., S.L., R.S., and U.C.; software, S.P., and U.C.; validation, S.L., R.S., and U.C.; formal analysis, M.C., Mi.C., A.K., S.P., and U.C.; investigation, G.W., R.Z., T.D., S.G., and R.M.S.; resources, S.L., R.S., and U.C.; data curation, S.P., G.W., R.Z., and R.S.; writing—original draft preparation, M.C., Mi.C., A.K., S.P., G.W., R.Z., T.D., S.G., R.M.S., S.L., R.S., and U.C.; writing—review and editing, S.P., S.L., R.S., and U.C.; visualization, S.P., S.L., R.S., and U.C.; supervision, S.L., R.S., and U.C.; project administration, U.C.; funding acquisition, S.L. All authors have read and agreed to the published version of the manuscript.

---